\documentclass[showpacs,eqsecnum,aps]{revtex4}

\usepackage[dvips]{graphicx,color}

\begin{document}
%\draft

\title{Three-Body Scattering at Intermediate Energies}

\author{H. Liu, Ch. Elster}
\affiliation{
Institute of Nuclear and Particle Physics,  and
Department of Physics and Astronomy, \\ Ohio University, Athens, OH 45701}

\author{ W. Gl\"ockle}
\affiliation{
Institute for Theoretical Physics II, Ruhr-University Bochum,
D-44780 Bochum, Germany.}

\vspace{10mm}

\date{\today}

%\maketitle

\begin{abstract}
The Faddeev equation for three-body scattering at arbitrary energies is
formulated in momentum space and directly solved in terms of momentum vectors 
without employing a partial wave
decomposition. In its simplest form the Faddeev equation for
identical bosons, which we are using, is a three-dimensional
integral equation in five variables, magnitudes of relative momenta and angles. 
This equation is solved 
through Pad$\acute{e}$ summation. Based on a Malfliet-Tjon-type potential, 
the numerical feasibility and stability of the algorithm for solving the
Faddeev equation is demonstrated.  Special attention is given
to  the selection of independent variables and the
treatment of three-body break-up singularities with a spline based method.
The elastic differential cross section, semi-exclusive d(N,N$^\prime$) cross sections and
total cross sections of both
elastic and breakup processes in the intermediate energy range up to about 1 GeV 
are calculated and the convergence of the multiple scattering series is investigated
in every case. In general a truncation in the first or second order in the two-body t-matrix is quite
insufficient.
\end{abstract}

\vspace{10mm}


\maketitle

%\newpage

%****************************************************************************

%\narrowtext

%****************************************************************************

\section{Introduction}

During the last two decades calculations of nucleon-deuteron scattering 
experienced large improvements and refinements. Here different techniques have been applied, Faddeev
calculations in configuration space \cite{benchpayne} and momentum space \cite{Cornelius}, and
variational calculations based on a hyperspherical harmonic expansion \cite{kievsky1,kievsky2}. 
It is fair to say that below about 200~MeV
projectile energy the momentum space Faddeev equations for three-nucleon scattering can now be solved
with high accuracy for the most modern two- and three-nucleon forces. A summary for these achievements
can be found in Ref.~\cite{wgphysrep,wgarticle,kuros,witala1,sauer}. 
The approach described there is based on using angular
momentum eigenstates for the two- and three-body systems. This partial wave decomposition replaces the
continuous angle variables by discrete orbital angular momentum quantum numbers, and thus reduces the
number of continuous variables to be discretized in a numerical treatment.
For low projectile energies the procedure of considering
orbital angular momentum components appears physically justified due
to arguments related to the centrifugal barrier and the short range of the nuclear force. 
However, the algebraic and algorithmic steps to be carried out in a partial wave
decomposition can be quite involved when solving the Faddeev
equations.  If one considers three-nucleon scattering at a few hundred
MeV projectile energy, the number of partial waves needed to achieve
convergence proliferates, and limitations with respect to
computational feasibility and accuracy are reached.  

It appears therefore natural to avoid a partial wave representation
completely and work directly with vector variables. This is common
practice in bound state calculations of few-nucleon systems based on
variational \cite{Arriaga95a} and Green's function Monte Carlo (GFMC)
methods \cite{Carlson87a,Carlson88a,Zabolitzki82a,Carlson99a}, which
are carried out in configuration space.

Our aim is to work directly with vector variables in the Faddeev
scheme in momentum space. In earlier work \cite{triton3d,3nfbound} 
we showed that the bound state Faddeev equation 
has a rather transparent structure when formulated with vector
variables compared to the coupled set of two-dimensional integral
equations obtained in a partial wave decomposed form.
Based on
Malfliet-Tjon type interactions for two- as well as three-body forces,
it was demonstrated that the numerical
solution of the bound state equation using vector variables is
straightforward and numerically very accurate.
As far as three nucleon scattering is concerned, the neutron-deuteron break-up process has
been successfully studied up to 500 MeV projectile energy
based on the first order term of the Faddeev equation using
realistic nucleon-nucleon forces \cite{Nd3d}. 

In this article we want to show that the full solution of the three-body
scattering equation can  be obtained in a straightforward manner
when employing vector variables, i.e. magnitudes of momenta and angles
between the momentum vectors.  As a simplification 
we neglect spin and iso-spin degrees of freedom and treat three-boson scattering. 
The interactions employed are of Yukawa type, and no separable approximations are
involved. The Faddeev equation for three identical bosons is
solved exactly as function of momentum vectors below and above the three-body 
break-up. 

This article is organized as follows. Section II reviews the 
Faddeev equation for three-body scattering in momentum space and discusses our
choice of momentum and angle variables for the unknown amplitude in the equation and its kernel.
In Section III we derive the amplitudes and cross sections for elastic scattering and break-up processes.
In addition we relate both via the optical theorem.
In Section IV we discuss the numerical methods necessary for solving the Faddeev equation, especially our
treatment of the singularities in the free three-body propagator. In addition, our numerical tests 
for the solution are shown and discussed.
In Section V we present calculations for elastic scattering and break-up processes in the intermediate
energy regime from 0.2 to 1~GeV. Our focus here is on the study of the importance of rescattering terms
as function of the projectile energy and the reaction considered.
We conclude in Section VI.

%****************************************************************************

\section{Faddeev Equations for Three Bosons in the Continuum}

Various presentations of three-body scattering in the Faddeev scheme are presented in the literature
\cite{wgbook,wgphysrep,wgarticle}. We solve the Faddeev equation for three identical particles in the
form
\begin{equation}
T|\phi\rangle=tP|\phi\rangle+tPG_{0}T|\phi\rangle. \label{eq:2.1}
\end{equation}
The driving term of this integral equation consists of the two-body t-matrix
$t$, the sum $P$ of a cyclic and anticyclic permutation of three particles, and the
initial state $|\phi\rangle = |\varphi_d {\bf q_0}\rangle$, composed of a two-body
bound state and the momentum eigenstate of the projectile particle.  
The kernel of of Eq.~(\ref{eq:2.1}) contains the free three-body propagator,
$G_{0}=(E-H_{0}+i\varepsilon)^{-1}$, 
where $E$ is the total energy in the center of mass frame.  

\noindent
The operator $T$ determines both, the full break-up amplitude
\begin{equation}
U_{0}=(1+P)T ,
\label{eq-U0}
\end{equation}
and the amplitude for elastic scattering
\begin{equation}
U=PG^{-1}_{0}+PT.
\label{eq-U}
\end{equation}
In this paper we focus on three identical bosons and use a momentum space representation.
For solving Eq.~(\ref{eq:2.1}), we introduce standard Jacobi momenta {\bf p}, the relative momentum
in the subsystem, and {\bf q}, the relative momentum of the spectator to the subsystem.
The momentum states are normalized according to $\langle {\bf p}'{\bf q}'|{\bf p} {\bf q}\rangle =
\delta^3 ({\bf p}'-{\bf p}) \delta^3 ({\bf q}' -{\bf q})$. 
Projecting  Eq.~(\ref{eq:2.1}) on to Jacobi momenta leads to \cite{scatter1}
\begin{eqnarray}
\langle{\mathbf{p}}{\mathbf{q}}|T|\varphi_{d}{\mathbf{q}}_{0}\rangle &=&
\varphi_{d}\left({\mathbf{q}}+\frac{1}{2}{\mathbf{q}}_{0}\right)
t_{s}\left({\mathbf{p}},\frac{1}{2}{\mathbf{q}}+{\mathbf{q}}_{0},E-\frac{3}{4m}q^{2}
\right)
\nonumber \\
&+&\int
d^{3}q''t_{s}\left({\mathbf{p}},\frac{1}{2}{\mathbf{q}}+{\mathbf{q}}'',
E-\frac{3}{4m}q^{2}\right)
\frac {\left\langle{\mathbf{q}}+\frac{1}{2}{\mathbf{q}}'',{\mathbf{q}}''
\left|T\right| \varphi_{d}{\mathbf{q}}_{0} \right\rangle}
{E-\frac{1}{m}(q^{2}+q''^{2}+{\mathbf{q}}\cdot{\mathbf{q}}'')+i\varepsilon}.
\label{eq:2.4}
\end{eqnarray}
Here
$t_{s}({\mathbf{p}}',{\mathbf{p}})=
t({\mathbf{p}},{\mathbf{p}}')+t(-{\mathbf{p}}',{\mathbf{p}})$
is the symmetrized $t$ matrix and $E$ is the total energy in the center of mass (c.m.) system
\begin{equation}
 E=E_{d}+\frac{3}{4m}q^{2}_{0} =E_{d}+\frac{2}{3}E_{lab}. \label{eq:2.5}
\end{equation}
We assume that the underlying force is a two-body force, generating $t$ via a two-body
Lippmann-Schwinger equation and supporting one bound state with energy $E_d$.
Thus, $t_{s}(z)$ has a pole at $z=E_d$. Since the transition operator $T$ in Eq.~(\ref{eq:2.4}) is
needed for all values of {\bf q}, one encounters this pole of $t_s$. 
Extracting the residue explicitly by defining
\begin{equation}
{t}_{s}({\mathbf{p}'},{\mathbf{p}},z) \equiv 
\frac{ {\hat t_{s}}({\mathbf{p}'},{\mathbf{p}},z)}{z-E_{d}}
\end{equation}
and similarly for $T$, Eq.~(\ref{eq:2.4}) can be rewritten as
\begin{eqnarray}
\langle{\mathbf{p}}{\mathbf{q}}|\hat{T}|\varphi_{d}{\mathbf{q}}_{0}\rangle &=&
\varphi_{d}\left({\mathbf{q}}+\frac{1}{2}{\mathbf{q}}_{0}\right)
\hat{t}_{s}\left( {\mathbf{p}},\frac{1}{2}{\mathbf{q}}+{\mathbf{q}}_{0},E-\frac{3}{4m}q^{2}
\right) \nonumber \\
&+&\int d^{3}q'' \frac {\hat{t}_{s}\left(
{\mathbf{p}},\frac{1}{2}{\mathbf{q}}+{\mathbf{q}}'',E-\frac{3}{4m}q^{2}\right)}
{E-\frac{1}{m}(q^{2}+q''^{2}+{\mathbf{q}}\cdot{\mathbf{q}}'')+i\varepsilon} \ 
\frac
{\left\langle{\mathbf{q}}+\frac{1}{2}{\mathbf{q}}'',{\mathbf{q}}''|\hat{T}|\varphi_{d}
{\mathbf{q}}_{0}\right\rangle}
{E-\frac{3}{4m}q''^{2}-E_{d}+i\varepsilon}. \label{Faddeev-hat-T}
\end{eqnarray}
This expression is the starting point for our numerical calculation of the transition amplitude
without employing an angular momentum decomposition.

The first important step for an explicit calculation is the selection of independent variables.
Since we ignore spin and iso-spin dependencies,  the matrix element 
$\langle {\bf p}{\bf q}|\hat{T}| \varphi_d {\bf q_0}\rangle$ is a scalar function of the
variables $\mathbf{p}$ and  $\mathbf{q}$ for a given projectile momentum
$\mathbf{q}_{0}$.
Thus one needs 5 variables to uniquely specify the geometry of the
three vectors $\mathbf{p}$, $\mathbf{q}$ and $\mathbf{q}_{0}$, which are shown in Fig.~1. 
Having in mind that
with three vectors one can span 2 planes, i.e. the $\mathbf{p}$-$\mathbf{q}_{0}$-plane and
the $\mathbf{q}$-$\mathbf{q}_{0}$-plane, a natural choice of independent variables is
\begin{equation}
p=|{\mathbf{p}}|,\  q=|{\mathbf{q}}|,\
x_{p}=\hat{{\mathbf{p}}}\cdot\hat{{\mathbf{q}}}_{0},\
x_{q}=\hat{{\mathbf{q}}}\cdot\hat{{\mathbf{q}}}_{0},\
x^{q_{0}}_{pq}=
\widehat{({\mathbf{q}}_{0}\times{\mathbf{q}})}\cdot\widehat{({\mathbf{q}}_{0}\times{\mathbf{p}})}.
\label{variables}
\end{equation}
The last variable, $x^{q_{0}}_{pq}$, is the angle between the two normal vectors of the 
$\mathbf{p}$-$\mathbf{q}_{0}$-plane and the $\mathbf{q}$-$\mathbf{q}_{0}$-plane,
which are explicitly given by
\begin{eqnarray}
\widehat{({\mathbf{q}}_{0}\times{\mathbf{p}})}&=&\frac{\hat{\mathbf{q}}_{0}\times\hat{\mathbf{p}}}
{\sqrt{1-(\hat{\mathbf{q}}_{0}\cdot\hat{\mathbf{p}})^{2}}},
\nonumber \\
\widehat{({\mathbf{q}}_{0}\times{\mathbf{q}})}&=&\frac{\hat{\mathbf{q}}_{0}\times\hat{\mathbf{q}}}
{\sqrt{1-(\hat{\mathbf{q}}_{0}\cdot\hat{\mathbf{q}})^{2}}}.
\label{normals}
\end{eqnarray}
It should be pointed out, that the angle between the vectors $\mathbf{q}$ and  $\mathbf{p}$,
$y_{pq}={\bf \hat p}\cdot {\bf \hat q}$, is {\bf not} an independent variable. In fact, if $x_p$ and
$x_q$ are given, the domain of $y_{pq}$ is bound by
\begin{equation}
x_{p}x_{q}-\sqrt{1-x^{2}_{p}}\sqrt{1-x^{2}_{q}}\leq y_{pq} \leq
x_{p}x_{q}+\sqrt{1-x^{2}_{p}}\sqrt{1-x^{2}_{q}},
\label{eq:2.10}
\end{equation}
thus not covering the entire interval [-1,1]. 
Using the explicit representation of the normal vectors and standard cross product identities, we
arrive at the following relation between $x^{q_{0}}_{pq}$ and $y_{pq}$, 
\begin{eqnarray}
x^{q_{0}}_{pq}&=&\frac{\hat{\mathbf{p}}\cdot\hat{\mathbf{q}}-
(\hat{\mathbf{q}}_{0}\cdot\hat{\mathbf{p}})(\hat{\mathbf{q}}_{0}\cdot\hat{\mathbf{q}})}
{\sqrt{1-(\hat{\mathbf{q}}_{0}\cdot\hat{\mathbf{p}})^{2}}\sqrt{1-(\hat{\mathbf{q}}_{0}\cdot\hat{\mathbf{q}})^{2}}}
\nonumber \\
&=&\frac{y_{pq}-x_{p}x_{q}}{\sqrt{1-x^{2}_{p}}\sqrt{1-x^{2}_{q}}},
\label{eq:2.11}
\end{eqnarray} 
or 
\begin{equation}
y_{pq}=x_{p}x_{q}+\sqrt{1-x^{2}_{p}}\sqrt{1-x^{2}_{q}}\
x^{q_{0}}_{pq}. \label{eq:2.12}
\end{equation}
For the special case where
$\hat{\mathbf{q}}_{0}$ is parallel to the $z$-axis ($q_{0}$-system) one can write
\begin{eqnarray}
y_{pq}=x_{p}x_{q}+\sqrt{1-x^{2}_{p}}\sqrt{1-x^{2}_{q}}\cos\varphi_{pq},
\label{eq:2.13}
\end{eqnarray}
where the $\varphi_{pq}$ is the difference of the azimuthal angles of
$\hat{\mathbf{p}}$ and $\hat{\mathbf{q}}$. However, the variable
$\cos\varphi_{pq}$, which was used erroneously in \cite{scatter1} as third angular variable, 
is not rotationally invariant. 

With the independent variables listed in Eq.~(\ref{variables}) the matrix element of 
 $\hat T$ is given as
\begin{equation}
\langle{\mathbf{p}}{\mathbf{q}}|\hat{T}|\varphi_{d}{\mathbf{q}}_{0}\rangle
 \equiv \hat{T}(p,x_{p}, x^{q_{0}}_{pq}, x_{q}, q; q_0).
\label{eq:2.14}
\end{equation}
Furthermore, ${\hat t}_s ({\bf p}',{\bf p},z)$ is also a scalar function, and thus can be written in
the form
\begin{equation}
{\hat t}_s ({\bf p}',{\bf p},z) = {\hat t}_s (p',p, {\hat {\bf p}}' \cdot {\hat {\bf p}},z).
\label{eq:2.15}
\end{equation}

The most intricate dependence appears under the integral in Eq.~(\ref{Faddeev-hat-T}) for the third
angular variable of the $\hat T$-amplitude.
According to Eq.~(\ref{eq:2.11}) it is given as
\begin{equation}
x^{q_0}_{(q+\frac{1}{2}q'')q''} \equiv \frac{ ({\widehat {{\bf q}+\frac{1}{2}{\bf q}''}}) 
\cdot {\bf {\hat q}}''
 - {\bf {\hat q}_0} \cdot (\widehat{ {\bf q} + \frac{1}{2} {\bf q}''}) \; {\bf {\hat q}_0} \cdot 
{\bf {\hat
q}}''} {\sqrt{1-\left({\bf {\hat q}_0} \cdot (\widehat{{\bf q} +\frac{1}{2} {\bf {\hat q}}''})\right)^2} 
 \; \; \sqrt{1-({\bf {\hat q}_0} \cdot {\bf {\hat q}}'')^2}}.
\label{eq:2.16}
\end{equation} 

In view of the break-up singularities of the first denominator in Eq.~(\ref{Faddeev-hat-T}) it is
mandatory to choose the coordinate system for the {\bf q}''-integration such that the z-axis points
parallel to the vector ${\hat {\bf q}}$. Then one obtains for Eq.~(\ref{eq:2.16}) 
\begin{equation}
x^{q_0}_{(q+\frac{1}{2}q'')q''} = \frac{ 
\frac{q x'' +\frac{1}{2} q''}{\sqrt{q^2+\frac{1}{4}q''^2 +qq''x''}}
 - x_{q+\frac{1}{2}q''} x_{q''}} 
{\sqrt{1- x^2_{q+\frac{1}{2}q''}} \;  \sqrt{1-x^2_{q''}}},
\label{eq:2.17}
\end{equation}
where
\begin{eqnarray}
x_{q''} &\equiv& {\bf {\hat q}''} \cdot {\bf {\hat q}_0}= x'' x_q + \sqrt{1-x''^2} \sqrt{1-x_q^2} 
\;\; \cos (\varphi'' -\varphi_{q_0}) \nonumber \\
x_{q+\frac{1}{2}q''}  &\equiv& (\widehat{{\bf q} +\frac{1}{2} {\bf q}''}) \cdot {\bf {\hat q}_0} =
 \frac{q x_q +\frac{1}{2} q'' x_{q''} }{\sqrt{q^2 +\frac{1}{4} q''^2 +q q''x''}}.
\label{eq:2.18}
\end{eqnarray}
Here $\varphi_{q_0}$ is the azimuthal angle of ${\bf {\hat q}_0}$ in the coordinate system chosen for the
{\bf q}''-integration.
These considerations lead to the explicit representation for the transition amplitude ${\hat T}$
\begin{eqnarray}
 \hat{T}(p,x_{p}, x^{q_{0}}_{pq}, x_{q}, q; q_0)
&=&\varphi_{d}\left(\sqrt{q^{2}+\frac{1}{4}q^{2}_{0}+qq_{0}x_{q}}\right)
 \label{eq:2.19}  \\
&\times&\hat{t}_{s} \left( p,
\sqrt{\frac{1}{4}q^{2}+q^{2}_{0}+qq_{0}x_{q}},
\frac{\frac{1}{2}qy_{pq}+q_{0}x_{p}}{\sqrt{\frac{1}{4}q^{2}+q^{2}_{0}+qq_{0}x_{q}}};
E-\frac{3}{4m}q^{2} \right) \nonumber \\
&+&\int^{\infty}_{0}dq''q''^{2}\int^{+1}_{-1}dx''\int^{2\pi}_{0}d\varphi''
\frac{1}{E-\frac{1}{m}(q^{2}+qq''x''+q''^{2})+i\varepsilon} \nonumber \\
&\times&\hat{t}_{s} \left( p,
\sqrt{\frac{1}{4}q^{2}+q''^{2}+qq''x''},
\frac{\frac{1}{2}qy_{pq}+q''y_{pq''}}{\sqrt{\frac{1}{4}q^{2}+q''^{2}+qq''x''}};
E-\frac{3}{4m}q^{2} \right) \nonumber \\
&\times& \frac{ {\hat T} \left(
\sqrt{q^{2}+\frac{1}{4}q''^{2}+qq''x''},
\frac{qx_{q}+\frac{1}{2}q''x_{q''}}{\sqrt{q^{2}+\frac{1}{4}q''^{2}+qq''x''}},
\frac{\frac{qx''+\frac{1}{2}q''}{\sqrt{q^{2}+\frac{1}{4}q''^{2}+qq''x''}}
-x_{q+\frac{1}{2}q''}x_{q''}}{\sqrt{1-x^{2}_{q+\frac{1}{2}q''}}\sqrt{1-x^{2}_{q''}}},
x_{q''},q'';q_0  \right) }
{E-\frac{3}{4m}q''^{2}-E_{d}+i\varepsilon}
\nonumber 
\end{eqnarray}
where in addition to Eq.~(\ref{variables}) and the related variables of Eq.~(\ref{eq:2.12}) and 
Eq.~(\ref{eq:2.18}) the following variables occur 
\begin{eqnarray}
q''&=&|\mathbf{q}''|, \nonumber \\
x''&=&\hat{\mathbf{q}}\cdot\hat{\mathbf{q}}'', \nonumber \\
y_{pq''}&=&\hat{\mathbf{p}}\cdot\hat{\mathbf{q}}''=
y_{pq}x''+\sqrt{1-x''^{2}}\sqrt{1-y^{2}_{pq}}\cos(\varphi_{p}-\varphi'').
\label{eq:2.20}
\end{eqnarray}
Like $\varphi_{q_{0}}$ in Eq.~(\ref{eq:2.18}), the angle $\varphi_p$ in Eq.~(\ref{eq:2.20}) is the azimuthal
angle of ${\hat {\bf p}}$ in the $q$-system (i.e. the system where the z-axis is parallel to ${\hat {\bf
q}}$). It remains to relate the angles $\varphi_p$ and $\varphi_{q_{0}}$ to the three angular variables 
$x_p$, $x_q$, and $x^{q_{0}}_{pq}$. As it is shown in Appendix~\ref{appendixa}, due to the
$\varphi''$-integration, only the knowledge of $\cos (\varphi_p -\varphi_{q_{0}})$ is required. Like $\cos
\varphi_{pq}$ in Eq.~(\ref{eq:2.13}) is equal to $x^{q_0}_{pq}$ in the $q_0$-system, so is 
$\cos (\varphi_p -\varphi_{q_{0}})$ equal to $x^q_{q_{0} p}$ in the $q$-system. Thus
\begin{eqnarray}
\cos(\varphi_{p}-\varphi_{q_{0}}) &=& x^{q}_{q_{0}p}
=\frac{\hat{\mathbf{q}}_{0}\cdot\hat{\mathbf{p}}-
(\hat{\mathbf{q}}\cdot\hat{\mathbf{q}}_{0})(\hat{\mathbf{q}}\cdot\hat{\mathbf{p}})}
{\sqrt{1-(\hat{\mathbf{q}}\cdot\hat{\mathbf{q}}_{0})^{2}}\sqrt{1-(\hat{\mathbf{p}}\cdot\hat{\mathbf{q}})^{2}}}
=\frac{x_{p}-x_{q} y_{pq}}{\sqrt{1-x^{2}_{q}} \; \sqrt{1-y^{2}_{pq}}}.
\label{eq:2.21}
\end{eqnarray}
Because of that difference $(\varphi_p -\varphi_{q_{0}})$,  one can choose 
$\varphi_{q_{0}}$ arbitrarily, e.g. zero.  Furthermore, $\cos \varphi_p$ and $\sin \varphi_p$ required in
Eq.~(\ref{eq:2.20}) are given in terms of $\cos (\varphi_p -\varphi_{q_{0}})$, as is shown in
Appendix~\ref{appendixa}. This completes the definition of all relevant variables in Eq.~(\ref{eq:2.19}).

\section{Amplitudes and Cross Sections for Elastic Scattering and Break-Up Processes}

The amplitude for elastic scattering is obtained by calculating the matrix element of the
operator $U$ 
given in Eq.~(\ref{eq-U}) as.
\begin{eqnarray}
\langle{\mathbf{q}}\varphi_{d}|U|{\mathbf{q}}_{0}\varphi_{d}\rangle
&=&2\varphi_{d}\left(\frac{1}{2}{\mathbf{q}}+{\mathbf{q}}_{0}\right)
\left(E-\frac{1}{m}(q^{2}+{\mathbf{q}}\cdot{\mathbf{q}}_{0}+q^{2}_{0})\right)
\varphi_{d}\left({\mathbf{q}}+\frac{1}{2}{\mathbf{q}}_{0}\right)  \nonumber \\
&+&2 \int d^{3}q''\varphi_{d}\left(\frac{1}{2}{\mathbf{q}}+{\mathbf{q}}''\right)
\frac{\left\langle {\mathbf{q}}+\frac{1}{2}{\mathbf{q}}'',{\mathbf{q}}''
|\hat{T}| {\mathbf{q}}_{0}\varphi_{d}\right\rangle}
{E-\frac{3}{4m}q''^{2}-E_{d}+i\varepsilon}.
\label{elastic-amp}
\end{eqnarray}

\noindent
The amplitude for the full break-up process according to Eq.~(\ref{eq-U0}), 
is given by
\begin{eqnarray}
\langle
{\mathbf{p}}{\mathbf{q}}|U_{0}|{\mathbf{q}}_{0}\varphi_{d}\rangle
&=&\frac{\left\langle {\mathbf{p}}{\mathbf{q}}|\hat{T}|{\mathbf{q}}_{0}\varphi_{d}\right\rangle}
{E-\frac{3}{4m}{\mathbf{q}}^{2}-E_{d}} +\frac{\left\langle
-\frac{1}{2}{\mathbf{p}}+\frac{3}{4}{\mathbf{q}},-{\mathbf{p}}-\frac{1}{2}{\mathbf{q}}|\hat{T}|
{\mathbf{q}}_{0}\varphi_{d}\right\rangle}{E-\frac{3}{4m}(-{\mathbf{p}}-\frac{1}{2}
{\mathbf{q}})^{2}-E_{d}} + \frac{\left\langle
-\frac{1}{2}{\mathbf{p}}-\frac{3}{4}{\mathbf{q}},+{\mathbf{p}}-\frac{1}{2}{\mathbf{q}}|\hat{T}|
{\mathbf{q}}_{0}\varphi_{d}\right\rangle}
{E-\frac{3}{4m}(+{\mathbf{p}}-\frac{1}{2}{\mathbf{q}})^{2}-E_{d}}
\nonumber \\
\label{breakup-amp}
\end{eqnarray}
The equation for the elastic operator $U$ follows from Eqs.~(\ref{eq:2.1}) and (\ref{eq-U}). It is
given as
\begin{equation}
U |\phi \rangle = P G_0^{-1} |\phi \rangle + P t G_0 U |\phi \rangle
\label{eq:3.3}
\end{equation}
Straightforward and well known steps \cite{wgphysrep}  based on this equation 
lead to the unitarity relation
\begin{eqnarray}
\langle\phi|U|\phi'\rangle^{*}
-\langle\phi'|U|\phi\rangle
&=&
\int d^{3}q \langle\phi_{q}|U|\phi'\rangle^{*} 2\pi i \delta(E-E_{d}-\frac{3}{4m}q^2)
\langle\phi_{q}|U|\phi\rangle \nonumber \\
&+&
\frac{1}{3}\int d^{3}p d^{3}q \langle\phi_{0}|U_{0}|\phi'\rangle^{*}
2\pi i \delta(E-\frac{p^2}{m}-\frac{3}{4m}q^2)\langle\phi_{0}|U_{0}|\phi\rangle.
\label{3nunitarity}
\end{eqnarray}
We want to point out that there is a misprint in Eq.~(202) of Ref.~\cite{wgphysrep}, the factor 1/3 is missing.

Using the variables defined in the previous section, and having in mind that for elastic
scattering $|{\bf q}|=|{\bf q_0}|$, the amplitude for elastic scattering  according to  Eq.~(\ref{elastic-amp})
can be expressed as
\begin{eqnarray}
\langle {\mathbf{q}}\varphi_{d}|U|{\mathbf{q}}_{0}\varphi_{d}\rangle &\equiv&  U (q_0,x_q) =
2\varphi^{2}_{d}\left(
q_{0}\sqrt{\frac{5}{4}+x_q}\right)\left(E-\frac{q^{2}_{0}}{m}(2+x_q)\right)
 \\
&+&2\int^{\infty}_{0}dq''q''^{2}\int^{+1}_{-1}dx''\int^{2\pi}_{0}d\varphi''
\frac{1}{{E-\frac{3}{4m}q''^{2}-E_{d}+i\varepsilon}} \nonumber \\
&\times&\varphi_{d}\left(
\sqrt{\frac{1}{4}q^{2}_{0}+q''^{2}+q_{0}q''y_{qq''}}\right) 
\nonumber \\
&\times& {\hat T} \left( 
\sqrt{q^{2}_{0}+\frac{1}{4}q''^{2}+q_{0}q''y_{qq''}},
\frac{q_{0}x_q+\frac{1}{2}q''y_{q_{0}q''}}{\sqrt{q^{2}_{0}
+\frac{1}{4}q''^{2}+q_{0}q''y_{qq''}}},
\frac{\frac{q_{0}y_{qq''}+\frac{1}{2}q''}{\sqrt{q^{2}_{0}
+\frac{1}{4}q''^{2}+q_{0}q''y_{qq''}}}-x_{\pi_{p}}x_{\pi_{q}}}{\sqrt{1-x^{2}_{\pi_{p}}}
\sqrt{1-x^{2}_{\pi_{q}}}}, y_{q_{0}q''}, q''; q_0 \right) \nonumber
\label{eq:3.4}
\end{eqnarray}
with
\begin{eqnarray}
y_{qq''}&=&\hat{\mathbf{q}}\cdot\hat{\mathbf{q}}'' \nonumber \\
y_{q_{0}q''}&=&\hat{\mathbf{q}}_{0}\cdot\hat{\mathbf{q}}'' = x_{\pi_{q}}  \nonumber \\
x_{\pi_{p}}&=&\frac{q_{0}x_q+\frac{1}{2}q''y_{q_{0}q''}}{\sqrt{q^{2}_{0}
+\frac{1}{4}q''^{2}+q_{0}q''y_{qq''}}}, 
\label{eq:3.5}
\end{eqnarray}
At this point, the choice of a specific coordinate system for the $\bf {q''}$-integration is still open. 
The angular variable $x_q={\hat {\bf q}}\cdot {\hat {\bf q_0}}$ 
represents the scattering angle. If the z-axis is chosen parallel to ${\bf {\hat q}_0}$, the angles
are
\begin{eqnarray}
y_{qq''}&=&x_q x''+\sqrt{1-x_q^{2}}\sqrt{1-x''^{2}}\cos(\varphi_{q}-\varphi''). \nonumber \\
y_{q_{0}q''}&=&x'' 
\label{eq:3.6}
\end{eqnarray}
While having the z-axis parallel to ${\bf {\hat q}_0}$ may be the intuitive choice to describe the
scattering with a given beam direction, we can  in principle also
choose the z-axis parallel to ${\bf {\hat q}}$. In that case the angels in Eq.~(\ref{eq:3.5}) are given
by
\begin{eqnarray}
y_{qq''}&=&x''  \nonumber \\
y_{q_{0}q''}&=&x_qx''+\sqrt{1-x_q^{2}}\sqrt{1-x''^{2}}\cos(\varphi_{q_{0}}-\varphi'').
\label{eq:3.7}
\end{eqnarray} 
The elastic cross section depends on the angle between the vectors ${\bf {\hat q}_0}$ and
${\bf {\hat q}}$, but not on the choice of z-axis. We use the possibility of calculating the
matrix elements of $U$ in the two different coordinate systems 
to test the quality of our numerical calculations.

The differential elastic cross section in the c.m. frame is given by
\begin{equation}
\frac{d\sigma_{el}}{d\Omega}=\left(\frac{2m}{3}\right)^{2}(2\pi)^{4}
|U(q_{0} x_q)|^{2},
\label{eq:3.8}
\end{equation}
and the corresponding total elastic cross section is
\begin{equation}
\sigma_{el}=\int d\Omega \frac{d\sigma_{el}}{d\Omega}
=\left(\frac{2m}{3}\right)^{2}(2\pi)^{5}\int^{+1}_{-1}dx|
|U(q_{0},x)|^{2}.
\label{eq:3.9}
\end{equation}

The full break-up amplitude is given in Eq.~(\ref{breakup-amp}). On the energy shell $p$ and $q$ are
constrained by $ p^2 +\frac{3}{4} q^2 = mE$.
As function of all five variables and the projectile momentum it reads
\begin{eqnarray}
U_0(p, x_{p}, x^{q_{0}}_{pq},x_{q},q,q_{0}) =
\frac{{\hat T} ( p, x_{p}, x^{q_{0}}_{pq},x_{q},q,q_0)}{E-\frac{3}{4m}q^{2}-E_{d}}+
\frac{{\hat T} ( p_{2}, x_{p_{2}}, x^{q_{0}}_{p_{2}q_{2}},
x_{q_{2}},q_{2},q_0)}{E-\frac{3}{4m}q_{2}^{2}-E_{d}} +
\frac{{\hat T} (p_{3}, x_{p_{3}},x^{q_{0}}_{p_{3}q_{3}}, x_{q_{3}},q_{3},q_0)}
{E-\frac{3}{4m}q_{3}^{2}-E_{d}}.
\label{eq:3.10}
\end{eqnarray}
Here the variables are defined as
\begin{eqnarray}
y_{pq}&=&x_{p}x_{q}+\sqrt{1-x^{2}_{p}}\sqrt{1-x^{2}_{q}}\
x^{q_{0}}_{pq}
\nonumber \\
p_{2}&=&|-\frac{1}{2}{\mathbf{p}}+\frac{3}{4}{\mathbf{q}}|=\frac{1}{2}\sqrt{p^{2}
+\frac{9}{4}q^{2}-3pqy_{pq}}
\nonumber \\
q_{2}&=&|-{\mathbf{p}}-\frac{1}{2}{\mathbf{q}}|=\sqrt{p^{2}+\frac{1}{4}q^{2}+pqy_{pq}}
\nonumber \\
p_{3}&=&|-\frac{1}{2}{\mathbf{p}}-\frac{3}{4}{\mathbf{q}}|=\frac{1}{2}\sqrt{p^{2}
+\frac{9}{4}q^{2}+3pqy_{pq}}
\nonumber \\
q_{3}&=&|+{\mathbf{p}}-\frac{1}{2}{\mathbf{q}}|=\sqrt{p^{2}+\frac{1}{4}q^{2}-pqy_{pq}}
\nonumber \\
x_{p_{2}}&=&\hat{\mathbf{p}}_{2}\cdot\hat{\mathbf{q}}_{0}=\frac{-\frac{1}{2}px_{p}
+\frac{3}{4}qx_{q}}{p_{2}}
\nonumber \\
x_{q_{2}}&=&\hat{\mathbf{q}}_{2}\cdot\hat{\mathbf{q}}_{0}=\frac{-px_{p}-
\frac{1}{2}qx_{q}}{q_{2}}
\nonumber \\
x_{p_{3}}&=&\hat{\mathbf{p}}_{3}\cdot\hat{\mathbf{q}}_{0}=\frac{-\frac{1}{2}px_{p}-
\frac{3}{4}qx_{q}}{p_{3}}
\nonumber \\
x_{q_{3}}&=&\hat{\mathbf{q}}_{3}\cdot\hat{\mathbf{q}}_{0}=\frac{+px_{p}-
\frac{1}{2}qx_{q}}{q_{3}}
\nonumber \\
x^{q_{0}}_{p_{2}q_{2}}
&=&\widehat{(\mathbf{q}_{0}\times\mathbf{p}_{2})}\cdot\widehat{(\mathbf{q}_{0}\times\mathbf{q}_{2})}=
\frac{\frac{\frac{1}{2}p^{2}-\frac{3}{8}q^{2}-\frac{1}{2}pqy_{pq}}{p_{2}q_{2}}-x_{p_{2}}x_{q_{2}}}
{\sqrt{1-x^{2}_{p_{2}}}\sqrt{1-x^{2}_{q_{2}}}}
\nonumber \\
x^{q_{0}}_{p_{3}q_{3}}
&=&\widehat{(\mathbf{q}_{0}\times\mathbf{p}_{3})}\cdot\widehat{(\mathbf{q}_{0}\times\mathbf{q}_{3})}=
\frac{\frac{-\frac{1}{2}p^{2}+\frac{3}{8}q^{2}-\frac{1}{2}pqy_{pq}}{p_{3}q_{3}}-x_{p_{3}}x_{q_{3}}}
{\sqrt{1-x^{2}_{p_{3}}}\sqrt{1-x^{2}_{q_{3}}}}
\label{eq:3.11} \end{eqnarray}
The five-fold differential break-up cross section is given in the c.m. frame
\begin{equation}
\frac{d^{5}\sigma_{br}}{d\Omega_{p}d\Omega_{q}dq}=
\frac{(2\pi)^{4}m^{2}}{3q_{0}}q^{2}\sqrt{mE-\frac{3}{4}q^{2}} \;\; 
|U_0(p, x_{p}, x^{q_{0}}_{pq},x_{q},q,q_{0}) |^{2}
\label{eq:3.12}
\end{equation}

It is convenient to calculate the total break-up cross section in the c.m. frame, since there
are no kinematic restrictions on the relative angles. For the explicit calculation we can 
make different choices of the z-axis, e.g. it can be  parallel to the direction ${\bf {\hat
q}_0}$ of the projectile, or parallel to either one of the Jacobi vectors ${\hat {\bf q}}$ and
${\hat {\bf p}}$. The different choices will obviously result in  different angular integrations. For
completeness we give all three choices here. This will be used as a highly non-trivial test of the
numerical results, as will be demonstrated in the next Section.
If the z-axis is parallel to ${\hat {\bf q_0}}$ we have
\begin{equation}
\int d\Omega_{p}
d\Omega_{q}=2\pi\int^{+1}_{-1}dx''_{p}\int^{+1}_{-1}dx''_{q}\int^{2\pi}_{0}d\varphi''_{pq}
\label{eq:3.13}
\end{equation}
with
\begin{equation}
x_{p} \rightarrow x''_{p}, \:\:\: x_{q} \rightarrow x''_{q},\:\:\:  
x^{q_{0}}_{pq} \rightarrow \cos\varphi''_{pq}.
\label{eq:3.14}
\end{equation}
If the z-axis is parallel to ${\hat {\bf q}}$, the angular integration becomes
\begin{equation}
\int d\Omega_{p}
d\Omega_{q}=2\pi\int^{+1}_{-1}dx''_{q}\int^{+1}_{-1}dy''_{pq}\int^{2\pi}_{0}d\varphi''_{pq_{0}}
\label{eq:3.15}
\end{equation}
with
\begin{eqnarray}
x_{p} &\rightarrow& x''_{q}y''_{pq}+\sqrt{1-x''^{2}_{q}}\sqrt{1-y''^{2}_{pq}}\cos\varphi''_{pq_{0}},
\nonumber \\
x_{q} &\rightarrow& x''_{q}, \nonumber \\
x^{q_{0}}_{pq} &\rightarrow&
\frac{y''_{pq}-\left(x''_{q}y''_{pq}+\sqrt{1-x''^{2}_{q}}\sqrt{1-y''^{2}_{pq}}\cos\varphi''_{pq_{0}}\right)x''_{q}}
{\sqrt{1-\left(x''_{q}y''_{pq}+\sqrt{1-x''^{2}_{q}}\sqrt{1-y''^{2}_{pq}}\cos\varphi''_{pq_{0}}\right)^{2}}
\sqrt{1-x''^{2}_{q}}}.
\label{eq:3.16}
\end{eqnarray}
Finally, if the z-axis is parallel to ${\hat {\bf p}}$, the angular integration is
\begin{equation}
\int d\Omega_{p}
d\Omega_{q}=2\pi\int^{+1}_{-1}dx''_{p}\int^{+1}_{-1}dy''_{pq}\int^{2\pi}_{0}d\varphi''_{qq_{0}}
\label{eq:3.17}
\end{equation}
with
\begin{eqnarray}
x_{p} &\rightarrow& x''_{p}, \nonumber \\
x_{q} &\rightarrow& x''_{p}y''_{pq}+\sqrt{1-x''^{2}_{p}}\sqrt{1-y''^{2}_{pq}}\cos\varphi''_{qq_{0}},
\nonumber \\
x^{q_{0}}_{pq} &\rightarrow&
\frac{y''_{pq}-\left(x''_{p}y''_{pq}+\sqrt{1-x''^{2}_{p}}\sqrt{1-y''^{2}_{pq}}\cos\varphi''_{qq_{0}}\right)x''_{p}}
{\sqrt{1-\left(x''_{p}y''_{pq}+\sqrt{1-x''^{2}_{p}}\sqrt{1-y''^{2}_{pq}}\cos\varphi''_{qq_{0}}\right)^{2}}
\sqrt{1-x''^{2}_{p}}}.
\label{eq:3.18}
\end{eqnarray}
Let us define a function ${\cal F}(p,q)$ as
\begin{equation}
{\cal F}(p,q) = \int d\Omega_{p}d\Omega_{q}|\langle\phi_{0}|U_{0}|\phi\rangle|^{2} ,
\label{eq:3.19}
\end{equation}
where the angle integrations over the break-up amplitude is carried out. This function should be
independent of the coordinate system in which the angle integrations are performed. 
We use this property to check our numerical calculations. 
This is a non-trivial test of our calculation, 
since especially at higher energies the transition amplitude ${\hat T}$
develops stronger angle dependencies, which challenge the accuracy of the multi-dimensional
interpolations. 

The angle integrated break-up cross section is given as
\begin{equation}
\frac{d\sigma_{br}}{dq}=\frac{1}{3}\frac{(2\pi)^{4}m^{2}}{3q_{0}}q^{2}
\sqrt{mE-\frac{3}{4}q^{2}} \;\; {\cal F} \left(\sqrt{mE-\frac{3}{4}q^{2}},q \right),
\label{eq:3.20}
\end{equation}
and the
total break-up cross section reads
\begin{equation}
\sigma_{br}=\frac{1}{3}\frac{(2\pi)^{4}m^{2}}{3q_{0}}
\int^{\sqrt{\frac{4mE}{3}}}_{0}dq q^{2}
\sqrt{mE-\frac{3}{4}q^{2}} \;\;
{\cal F} \left(\sqrt{mE-\frac{3}{4}q^{2}},q \right).
\label{eq:3.21}
\end{equation}
Using now the unitarity relation from Eq.~(\ref{3nunitarity}) the optical theorem gives
\begin{equation}
\sigma_{tot}=  \sigma_{el} + \sigma_{br} = -\frac{4m(2\pi)^{3}}{3q_{0}}
{\mathrm{Im}} U \left( q_{0}, 1 \right).
\label{eq:3.22}
\end{equation}
For later use we also mention the semi-exclusive cross section, where only one particle is 
detected in the break-up process
\begin{equation}
\frac{d\sigma}{d{\Omega_q} dq} = (2\pi)^4 \frac{m^2}{3 q_0} p q^2 \int d{\hat p} 
|U_0(p,x_p,x^{q_{0}}_{pq},x_q,q,q_0)|^2
\label{eq:3.23}
\end{equation}

\section{Numerical Methods}

The fully off-shell two-body t-matrix $t({\bf p'},{\bf p},z)$ 
is solved directly from the Lippmann-Schwinger equation
 as function of its vector variables \cite{t-matrix} for the off-shell energies $E -\frac{3}{4m}q^2$ as
required by Eq.~(\ref{eq:2.4}). The Faddeev equation is iterated, generating the multiple scattering
series, which is then summed by the Pad$\acute{e}$ method \cite{pade-1,hueber}. We use it in the form of
a continued fraction expansion as layed out in Ref.~\cite{wgbook}. 

The first integration to be performed in solving Eq.~(\ref{eq:2.19}) by iteration is the integration
over the azimuthal angle $\varphi''$. This leads to a function of variables, $q''$ and $x''$ and requires
interpolation in the second and third arguments of $t_s$ and the first four arguments of ${\hat T}$.
Our spline interpolations are based on the cubic Hermite splines given in \cite{hueber97-1}.

Let $F(q'',x'')$ be the resulting function in each step of the iteration. Clearly, it depends in addition
on the fixed variables $p$, $q$, $x_p$, $x_q$, and $x^{q_{0}}_{pq}$, which are omitted for clarity.
Then the next task is performing the remaining two singular integrations, 
\begin{equation}
I=\int^{\infty}_{0}dq''q''^{2}\int^{+1}_{-1}dx''\frac{F(q'',x'')}
{\left(E-\frac{1}{m}(q^{2}+q''^{2}+qq''x'')+i\varepsilon\right)
\left(E-\frac{3}{4m}q''^{2}-E_{d}+i\varepsilon\right)} 
\label{eq:4.1}
\end{equation}
If the c.m. energy $E$ is below the three-body break-up threshold, only the second denominator is
singular, and the simple pole can be treated by standard subtraction methods.

The intricate problem arises above the three-body break-up threshold, when in addition the first
denominator can vanish. It has the form

\begin{equation}
\frac{1}{E-\frac{1}{m}(q^{2}+qq''x''+q''^{2})+i\varepsilon}
=\frac{\frac{m}{qq''}}{x_{0}-x''+i\varepsilon}
\label{eq:4.2}
\end{equation}
with
\begin{equation}
x_{0}=\frac{mE-q^{2}-q''^{2}}{qq''}.
\label{eq:4.3}
\end{equation}
For $|x_{0}|\le 1$ a so-called moving singularity arises in the $q''$-$x''$-integration, since
$x_0$ depends on $q$. The direct treatment of those moving singularities using real variables has been
discussed in the literature \cite{Cornelius}.  We briefly review the appearance of these singularities
in form of logarithms, since we introduce a new quasi-analytic integration based on Spline functions.
The condition  $|x_{0}| = 1$ leads to the pole positions
\begin{equation}
q''=\pm\frac{q}{2}\pm\sqrt{mE-\frac{3}{4}q^{2}}, 
\label{eq:4.4}
\end{equation}
and one arrives at the well known shape in the $q$-$q''$-plane for $|x_{0}|\le 1$ shown in Fig.~2.
This region is bounded by

\begin{equation}
q_{+}=\frac{q}{2}+\sqrt{Q^{2}_{0}-\frac{3}{4}q^{2}} 
\label{eq:4.5}
\end{equation}
and
\begin{eqnarray}
q_{-}=\left\{
\begin{array}{cc}
 -\frac{q}{2}+ \sqrt{Q^{2}_{0}-\frac{3}{4}q^{2}} & \:\: q<Q_{0} \\
 +\frac{q}{2}- \sqrt{Q^{2}_{0}-\frac{3}{4}q^{2}} & \:\: q>Q_{0}
\end{array}\right.
\label{eq:4.6}
\end{eqnarray}
where $Q_{0}=\sqrt{mE}$.
Apparently there is no singularity if $q > q_{max} \equiv \sqrt{\frac{4m}{3}E}$. We distinguish four
cases, $q=0$, $0<q<q_{max}$, $q=q_{max}$, and $q > q_{max}$. The case $q=0$ reduces to a simple
subtraction and will not be discussed.
For $0<q<q_{max}$ we only consider the part of the $q''$-integration which contains the moving
singularities. It has  the schematic form
\begin{equation}
I'= \int^{q_{max}}_{0}dq''
\int^{+1}_{-1}dx'' \frac{f(q'',x'')}{x_{0}-x''+i\varepsilon}.
\label{eq:4.7}
\end{equation}
The first step is to perform a subtraction of the pole, which we carry out in the entire square
$0 \leq q,q'' \leq q_{max}$ by defining
\begin{eqnarray}
\hat{f} (q'',x_{0})=\left\{
\begin{array}{cc}
 f(q'',x_{0}): & |x_{0}| \le 1 \\
 f(q'',\frac{x_{0}}{|x_{0}|}): &  |x_{0}|>1
\end{array}\right.
\label{eq:4.8}
\end{eqnarray}
We obtain
\begin{eqnarray}
I'&=& 
 \int^{q_{max}}_{0}dq''\int^{+1}_{-1}dx''
\frac{f(q'',x'')-\hat{f}(q'',x_{0})}{x_{0}-x''}\nonumber \\
&+&
 \int^{q_{max}}_{0}dq''\hat{f}(q'',x_{0})
\ln\left|\frac{1+x_{0}}{1-x_{0}}\right|
 -i\pi\int^{q_{max}}_{0}dq''\Theta(1-|x_{0}|)
\hat{f}(q'',x_{0}).
\label{eq:4.9}
\end{eqnarray}
Now we define $q_{-}=\frac{q}{2}-\sqrt{Q^{2}_{0}-\frac{3}{4}q^{2}}$ and obtain
\begin{equation}
\ln\left|\frac{1+x_{0}}{1-x_{0}}\right| =
\left(-\frac{q_{-}}{|q_{-}|}\ln|q''+|q_{-}||-\ln|q''+q_{+}|\right)+
\left(+\frac{q_{-}}{|q_{-}|}\ln|q''-|q_{-}||+\ln|q''-q_{+}|\right).
\label{eq:4.10}
\end{equation}
This leads to the well separated part of the integral which contains the logarithmic singularity
\begin{equation}
\int^{q_{max}}_{0}dq''\hat{f}(q'',x_{0})
\left(+\frac{q_{-}}{|q_{-}|}\ln|q''-|q_{-}||+\ln|q''-q_{+}|\right).
\label{eq:4.11}
\end{equation}
It is here that we introduce the new technique that relies on cubic spline integration.

We divide the range of integration $[0,q_{max}]$ into intervals bounded
 by a set of grid points $q_i$. The set of grid points is supposed to be dense enough to
interpolate the function $ {\hat f}(q'',x_0) \equiv f( q'')$ sufficiently well by cubic
 Hermitean splines \cite{hueber97-1}. In Ref.~\cite{hueber97-1} a detailed presentation of these
spline functions is given specific for our use.  For the convenience
of the reader we now switch to the notation of \cite{hueber97-1} and denote the end
 points of the $i$-th
interval  by $x_1$ and $x_2$ and the two adjacent grid points to the left and
right of the $i$-th interval by $x_0$ and $x_3$, respectively.
Then as detailed in \cite{hueber97-1} the
 interpolating function in
the $i$-th interval (replacing $f(q'') \equiv f(x)$) can be written  as
\begin{equation}
f_i(x) = \sum_{j=0}^3 S_j(x) f(x_j),
\label{eq:4.12}
\end{equation}
where the modified spline functions are
\begin{eqnarray}
S_{0}(x)&=&-\phi_{3}(x)\frac{x_{2}-x_{1}}{x_{1}-x_{0}}\frac{1}{x_{2}-x_{0}},
\nonumber \\
S_{1}(x)&=&\phi_{1}(x)+\phi_{3}\left(
\frac{x_{2}-x_{1}}{x_{1}-x_{0}}-\frac{x_{1}-x_{0}}{x_{2}-x_{1}}\right)\frac{1}{x_{2}-x_{0}}
-\phi_{4}(x)\frac{x_{3}-x_{2}}{x_{2}-x_{1}}\frac{1}{x_{3}-x_{1}},
\nonumber \\
S_{2}(x)&=&\phi_{2}(x)+\phi_{3}\frac{x_{1}-x_{0}}{x_{2}-x_{1}}\frac{1}{x_{2}-x_{0}}
+\phi_{4}(x)\left(
\frac{x_{3}-x_{2}}{x_{2}-x_{1}}-\frac{x_{2}-x_{1}}{x_{3}-x_{2}}\right)\frac{1}{x_{3}-x_{1}},
\nonumber \\
S_{3}(x)&=&\phi_{4}(x)\frac{x_{2}-x_{1}}{x_{3}-x_{2}}\frac{1}{x_{3}-x_{1}}.
\label{eq:4.13}
\end{eqnarray}
with
\begin{eqnarray}
\phi_{1}(x)&=&\frac{(x_{2}-x)^{2}}{(x_{2}-x_{1})^{3}}\left[
(x_{2}-x_{1})+2(x-x_{1})\right], \nonumber \\
\phi_{2}(x)&=&\frac{(x_{1}-x)^{2}}{(x_{2}-x_{1})^{3}}\left[
(x_{2}-x_{1})+2(x_{2}-x)\right], \nonumber \\
\phi_{3}(x)&=&\frac{(x-x_{1})(x_{2}-x)^{2}}{(x_{2}-x_{1})^{2}},
\nonumber \\
\phi_{4}(x)&=&\frac{(x-x_{1})^{2}(x-x_{2})}{(x_{2}-x_{1})^{2}}.
\label{eq:4.14}
\end{eqnarray}
Therefore, in view of Eqs.~(\ref{eq:4.11})- (\ref{eq:4.14}), the following integrals occur for $i=1 \cdots 4$
\begin{eqnarray}
\overline{\phi}_{1}&=&\int^{x_{i+1}}_{x_{i}}\phi_{1}\ln|x-q|dx, \nonumber \\
\overline{\phi}_{2}&=&\int^{x_{i+1}}_{x_{i}}\phi_{2}\ln|x-q|dx, \nonumber \\
\overline{\phi}_{3}&=&\int^{x_{i+1}}_{x_{i}}\phi_{3}\ln|x-q|dx, \nonumber \\
\overline{\phi}_{4}&=&\int^{x_{i+1}}_{x_{i}}\phi_{4}\ln|x-q|dx.
\label{eq:4.15}
\end{eqnarray}
with $q={ |q_-|,q_+}$.  Consequently the five different cases $q<x_{i}<x_{i+1}$, $q=x_{i}<x_{i+1}$,
$x_{i}<q<x_{i+1}$, $x_{i}<q=x_{i+1}$, and $x_{i}<x_{i+1}<q$ occur. Since the functions
$\phi_i (x)$ are cubic polynomials, the integrals in Eq.~(\ref{eq:4.15}) can be performed analytically.
We leave the explicit calculation to interested practitioner and refer to \cite{hangthesis} for
a detailed presentation.
According to our experience that manner to integrate the moving logarithmic
 singularities is a very good alternative  to the more common subtraction method
\cite{Cornelius}.

Finally, for $q=q_{max}$ we also apply the subtraction over the extended region $0\leq q'' \leq q_{max}$. 
In that case  $q_{+}=q_{-}=\frac{q_{max}}{2}$  and when $q''=\frac{q_{max}}{2}$ then $x_{0}=-1$.
Analogous steps lead to that part of the integral, which contains the logarithmic singularity,
\begin{equation}
\int^{q_{max}}_{0}dq''\hat{f}(q'',-1)\ln\left|q''-\frac{q_{max}}{2}\right|,
\label{eq:4.16}
\end{equation}
which is again evaluated by spline based integration.

In order to test the correctness as well as the accuracy of our calculations we carried out
a variety of numerical tests. Unfortunately we could not
compare to work by other groups, since to the best of our knowledge
 no comparable work at higher energies exists.

Apart from the projectile momentum $q_0$, the amplitude ${\hat T}$ of Eq.~(\ref{eq:2.19})  
depends on five variables $p$, $x_p$, $x_{pq}^{q_{0}}$,$x_q$,and $q$. In addition, there are the
integration variables $q''$, $x''$ and $\varphi''$.
All calculations listed are based on the Malfliet-Tjon type potential which is explicitly
given in the next section.
The fully off-shell two-body $t$-matrix, $t(p',p,x,\varepsilon)$, is obtained for each
fixed energy on a symmetric momentum grid with 60 $p$ ($p'$) points and 40 $x$ points.
 Since the momentum
region which contributes to a solution of the two-body $t$-matrix is
quite different from the region of importance in a three-body
calculation, we map our solution for $t_{\rm s}$ onto the momentum grid
relevant for the three-body transition amplitude. This is done by
applying the Lippmann-Schwinger equation repeatedly. The $t$-matrix
$t_{\rm s}(p',p,x, \varepsilon)$ is obtained at energies $\varepsilon = E -
\frac{3}{4m} q^2$, exactly at the $q$ values needed in the three-body
transition amplitude of Eq.~(\ref{eq:2.19}).

In carrying out our calculations, it turns out that there are essentially two separate
issues governing the quality of the results.  The first is the angle dependence of
the transition amplitude of  Eq.~(\ref{eq:2.19}). It is to be expected that the
angle dependence is weak at low energies and increases with higher energies,
reflecting the need to include more and more partial waves at higher energies in a
partial wave based calculations.  As example we list in Table~\ref{table-2} the
elastic and break-up total cross sections together with the total cross section
extracted from the imaginary part of $U$ in forward direction, Eq.~(\ref{eq:3.22}).
At 0.01~GeV 12 points for all angles are clearly sufficient, whereas at 0.1~GeV 
this is not so. Table~\ref{table-2} lists the elastic, break-up and total cross
section as function of the angle variables, and we see that one needs at least 16 
points for all angles.
At 0.5~GeV we find that the biggest angular
dependence occurs in $x_q$ and $x"$, and the least dependence in the azimuthal
angle $\varphi"$ and the angle $x^{q_0}_{pq}$, 
and take this into account in our choice of angle points.  

The other issue is the quality of the calculation
in the singular regime, i.e. in the integration region bounded by $q_{max}$ in
Fig.~\ref{fig2}. We divide the integration grid for $q"$ into the intervals
$ (0,q_{max}) \cup (q_{max},{\bar q})$, where
$q_{max}=\sqrt{\frac{4m}{3}E}$ and ${\bar q}=$20~fm$^{-1}$. The inteval boundaries
$0$ and $q_{max}$ are handled explicitly.  As the energy 
increases, $q_{max}$ increases, and we need to take this into account by changing the
distribution of the $q"$ points as function of energy within the different
$q''$-intervals, i.e. put more points into the interval $(0,q_{max})$ and less into 
$(q_{max},{\bar q})$. From the number of points in $(0,q_{max})$ one can define an average 
point
distance $\Delta_q \equiv q_{max}/({\rm number \ of \ points \ in}(0,q_{max}))$ 
in this interval.
In Fig.~\ref{fig3a} we show the dependence of the calculation on $\Delta_q$ by using
the percent error $\delta_{opt} = |\sigma_{opt} -\sigma_{el} -\sigma_{br}|/\sigma_{opt}
\times 100$ in the fulfillment of the optical theorem as quality measure.
At 0.01~GeV it is quite easy to make the average point distance very small
in the interval $(0,q_{max})$, since $q_{max}$ is only 75~MeV. The top panel of
Fig.~\ref{fig3a} shows that the percent error $\delta_{opt}$ drops linearly below
0.1\% and flattens out at $\Delta_q$= 3.5~MeV where most likely errors in the 
interpolation start to play a role. 
At projectile energy 0.1~GeV $q_{max}$ is already 284~MeV, and $\Delta_q$ is
naturally much larger with a reasonable number of $q''$ points. The dependence of
the $\delta_{opt}$ for 0.1~GeV on $\Delta_q$ is shown in the middle panel of
Fig.~\ref{fig3a} for two different cases. An angular grid size of 12 points is
indicated by the open squares, one of 24 points by crosses. A comparison of the
calculations shows at $\Delta_q$ = 15~MeV the calculation with 12 angular points
can not be improved any further, the calculations start to oscillate for smaller
$\Delta_q$, accidentally giving a very good agreement at $\Delta_q$ = 13~MeV.
Increasing the number of angle points to 24 shows a further linear decrease in the
error (cross symbols) into the 1\% region of $\delta_{opt}$ at a $\Delta_q$ = 9~MeV. 
This is consistent with the findings shown in the top panel.  
We continue to study the dependence of $\Delta_q$ at 0.5~GeV, where $q_{max}$ =
644~MeV. Here we immediately use 24 angle points as suggested from
Table~\ref{table-2}. A total of 30 integration points in $(0,q_{max})$ leads to
$\Delta_q$ = 22~MeV and $\delta_{opt} \approx$ 10\%, which is consistent with
values in the middle panel.  The 10\% error is also consistent with 
the value for the total break-up cross section in the last row of
Table~\ref{table-2}, which indicates that $\sigma_{br}$ is not yet converged.
From the systematics at the different energies 
shown in Fig.~\ref{fig3a} we can extrapolate on the $\Delta_q$ needed to
reduce the error in the optical theorem. Due to computer time limitations we have
not pushed this any further.
 
In Table~\ref{table-2} the total elastic, the total breakup as well as the total cross section 
evaluated according to Eq.~(\ref{eq:3.22}) via
the optical theorem are given and shown  as function of various sets of grid points. 
The momentum grids for
$p$ and $q$ are discretized with 49 points each. The integration variable $q''$
plays the same role as $q$,
and is therefore also discretized with 49 points distributed over the
intervals $(0,q_{max})\cup (q_{max},{\bar q})$ in an energy dependent way according
to the insights described above.
The values given in last row of each energy correspond to the points at 
the smallest $\Delta_q$ in Fig.~\ref{fig3a}. 

A nontrivial test for the quality accuracy of our calculation is the numerical verification of the
optical theorem Eq.~(\ref{eq:3.22}). Our results are given for selected energies in
Table~\ref{table-3}.
Here we show two sets of cross sections, distinguished by the superscripts $q_0$ and $q$ 
for the total and elastic cross section. 
The
superscripts indicate that the calculation is carried out by choosing the z-axis either parallel to 
${\bf {\hat q_0}}$  or to ${\bf {\hat q}}$. Performing the calculation with two different 
choices of the
z-axis is a nontrivial test for our choice of independent variables as well as for the entire
calculation. The total break-up cross section is also calculated in a coordinate
system in which the z-axis is parallel to ${\bf {\hat p}}$, indicated by
$\sigma^p_{br}$.
The calculations are based on the largest grids given in Table~\ref{table-2} and show
a very good agreement of the results obtained in the different coordinate systems.
This indicates the numerical rotational invariance of our calculations.

On top of convergence tests for the Pad$\acute{e}$ summation, we insert 
the resulting amplitude ${\hat T}$ 
again into the integral of Eq.~(\ref{eq:2.19}), leading to a second amplitude  ${\hat T}'$. 
Both amplitudes should be identical
within our numerical errors. This is checked by evaluating the  cross
sections again using the second amplitude. We document the results in Table~\ref{table-4}  for the
differential elastic cross section at selected angles. The table shows excellent agreement
of the two values of the cross section

Finally, another  highly nontrivial test of our calculation is the independence of the cross sections
from the arbitrary
angle $\varphi_{q_{0}}$ and the sign of $\sin(\varphi_p -\varphi_{q_{0}})$. 
This is documented in Tables~\ref{table-5}  and \ref{table-6} 
for the energy $E$=3~MeV. In order the check the rotational invariance numerically, the
calculations are carried out in two different coordinate systems, one where ${\bf q_0}$ is
parallel the $z$-axis, and one where ${\bf q}$ is parallel to $z$. Both tables show excellent
agreement for the cross sections, and thus we conclude that our choice of variables is correct.

\section{Scattering Calculations at Intermediate Energies}

Through we neglect spin and iso-spin degrees of freedom and stay in a strictly nonrelativistic
framework, we nevertheless can provide first qualitative insights for various cross sections in
three-body scattering in the intermediate energy regime which we define from 200~MeV to 1~GeV projectile
energy. The focus of our investigations will be the question which orders of rescattering in the two-body
$t$-matrix are needed to come close to the exact result, namely the solution of the Faddeev equation.

As a model two-body interaction we choose a superposition of two Yukawa interactions of
 Malfliet-Tjon type \cite{malfliet69-1}
\begin{eqnarray}
V({\mathbf{p}}', {\mathbf{p}})=\frac{1}{2\pi^{2}}
\left(
\frac{V_{R}}{({\mathbf{p}}'-{\mathbf{p}})^{2}+\mu^{2}_{R}} - 
\frac{V_{A}}{({\mathbf{p}}'-{\mathbf{p}})^{2}+\mu^{2}_{A}} 
\right).
\end{eqnarray}
The parameters are given in Table~\ref{table-7}, and fitted such that the potential supports a 
two-body bound state, the `deuteron', at -2.23~MeV. As first result we show in 
Fig.~\ref{fig3} the total cross section together with the total elastic and total break-up cross section
as function of the laboratory projectile energy. In addition, the total cross section
is also evaluated via the optical theorem as a test of our numerics. This duplicates the
information already given in Table~III.
We see that the optical theorem is quite well fulfilled. The figure shows that at roughly
 1~GeV the total elastic and total
break-up cross section become equal in magnitude in our model. 

Next we show in Fig.~4 the angular distribution in elastic scattering for a set of selected
energies. In addition to the exact Faddeev result,  the cross sections are evaluated in 
first order in the two-body $t$-matrix, 
second order in $t$, third order in $t$, and 4th order in $t$, and displayed. First we notice that
with increasing energy the cross section in forward direction decreases. 
Furthermore, for all energies
shown, the first rescattering (2nd order in $t$) always increases the cross section, and subsequent
rescatterings lower it again. 
As expected, for the lowest energy, 0.2~GeV, rescattering terms of higher order are
important, and even the 4th order is not yet close to the full result. The same is true for 0.8~GeV.
We notice, that even at 1~GeV two rescattering terms (3rd order in $t$)
are necessary to come into the vicinity of the final result. The same is true for 0.5~GeV. 

In view of the standard `$t$-$\rho$' impulse approximation for the optical potential in
nucleon-nucleus scattering employed at intermediate energies\cite{pNscat},
 it is interesting to notice that the first order result in  $t$ in our
model study is quite insufficient. Even at energies larger than 0.5~GeV rescattering corrections up to the 
3rd order are required to come close to the exact result for small scattering angles. 
Therefore it seems to be likely that the first order impulse
approximation in nucleon-nucleus scattering is insufficient.

In case of inelastic processes we first regard the semi-exclusive reaction d(N,N$^\prime$) where
only one nucleon is detected. We choose three different laboratory energies, 200~MeV, 500~MeV, and 1~GeV
and show the inclusive cross section as a
few selected angles for the detected nucleon. The results are shown in Figs.~\ref{fig5} through
\ref{fig9}.

At 0.2~GeV the semi-exclusive cross section is given in Fig.~\ref{fig5} for the emission angle 24$^o$ and in
Fig.~\ref{fig6} for the emission angle 39$^o$. 
Both figures show in the upper panel the entire energy range of the emitted particle.
Since the cross section varies by two orders of magnitude, we display it in a logarithmic scale. In order
to better flash out the peak structures, the two lower panels show the high and low energies of the
emitted particle in a linear scale. Together with the full solution of the Faddeev equation (solid line)
we display the sums of the lowest orders of the multiple scattering series as indicated in the
figure. The peak at the highest energy of the emitted particle is the socalled final state
interaction (FSI) peak, which
only develops when rescattering terms are taken into account. This peak is a general feature
of semi-exclusive scattering and is present for all energies. The next peak is the  quasi-free (QFS) peak, and
one sees that at both angles one needs at least
rescattering of 4th order to come close to the full result. However, in contrast to the smaller angle, at
the larger angle, 39$^o$ in Fig.~\ref{fig6},  
the first order result for the larger energies is surprisingly close to the full
solution, though the multiple scattering series is by no means converged, 
as the following higher orders indicate. We also observe that the QFS peak moves to lower energies
of the emitted particle with increasing emission angle.
At both angles the very low energies of the emitted particle
exhibit a strong peak in first order, which is considerably lowered by the first rescattering. 
Here the calculation up to 3rd order in the multiple scattering series seems already sufficient.

For 0.5~GeV incident energy the semi-exclusive cross section is given in Fig.~\ref{fig7} for the emission
angle 24$^o$ and in Fig.~\ref{fig8} for the emission angle 36$^o$.
We again see three peaks along 
along the energy axis of the detected nucleon, the FSI and QFS peaks as well as the peak at the 
extreme low energy of the emitted particle.
Again we see that the results based on first and second order in $t$ alone
are quite insufficient and higher order rescatterings can not be neglected.
It is also interesting to observe that at 24$^o$, Fig.~\ref{fig7},  the third and higher order 
rescattering terms  
shift the peak to higher  energies, whereas at the larger angle of 36$^o$  
the peak positions  of the various orders coincide more or less and agree with the peak
position of the full calculation. Again, for the peak for the very low energies of the emitted particle,
the 3rd order calculation agrees already quite well with the full result.

At 1~GeV the situation is similar. As examples we have selected two angles,
18$^o$ displayed in Fig.~\ref{fig9}  and 30$^o$ displayed in Fig.~\ref{fig10}.
For the small emission angle the second rescattering shifts the QFS peak towards higher energies, at the
larger angle this is not the case.
Our studies indicate that this is a general phenomemon occurring at all energies
under consideration. There exists a critical maximum energy $E_1^{max}$ of the emitted
particle, corresponding to a specific emission angle, at which such a shift in the
QFS peak through higher order rescattering terms can occur. At 0.5~GeV projectile
energy this maximum
energy is 0.44~GeV, at 1~GeV it is 0.88~GeV. If one considers the ratio 
$\frac{E_1^{max}}{E_{lab}}$, then one finds for both cases that if this ratio is 
larger than 0.8, the QSF peak is shifted by higher order rescattering terms. This
could be interpreted as an interference between the QFS and the FSI mechanisms.
If this ratio is smaller than 0.8, then the FSI peak is small and higher orders in
the multiple scattering series do not change the position of the QFS peak. 
 In addition, it  seems that at the larger angle (Fig.~\ref{fig10}) the multiple
scattering series converges a little faster with respect to the higher orders compared to the smaller
angle (Fig.~\ref{fig9}). The final result for the peak at the very low energy of the emitted particle
is as before reached with two rescattering contributions. It is remarkable that for the energies between
about 200 and 500~MeV of the emitted particle the first rescattering contributes almost an order of
magnitude to the cross section.

 We can make a first contact to calculations based on realistic nucleon-nucleon (NN) forces. 
In Ref.~\cite{Nd3d}  the
semi-exclusive process $d(p,n)$ has been determined in first order in $t$ based on the
NN potentials AV18 \cite{av18} and Bonn-B \cite{bonnb}. 
In the upper panel of Fig.~\ref{fig11} we compare our first order calculation at projectile
laboratory energy 495~MeV and 18$^o$ emission angle with the first order calculations from
Ref.~\cite{Nd3d} based on the two realistic potentials. The position of the peak is only determined
through kinematics, thus the peak position coincides for all three calculations. Though our model
calculation refers to bosons and the potential contains only the crude features of a central short range
repulsion and intermediate range attraction, the magnitudes of the cross sections differ only by
roughly 20\%. In the lower panel we show the contributions of the first orders of the
multiple scattering series successively summed together with the exact solution of the Faddeev equation
for our model. At this angle and energy the contribution of the first rescattering (2nd-order in the
multiple scattering series) is quite weak, the contributions of the next two orders are large and lower
the size of the peak. At the very high energies of the emitted particle, the 4th order in the multiple
scattering series is still not yet close to the exact result. Therefore we conjecture that at this energy
calculations with realistic forces will also require higher order rescattering contributions.

Finally we comment on a recently measured and analyzed reaction $pd \rightarrow (pp)n$ at high
momentum transfer \cite{komarov,haidenbauer}. In this experiment the break-up configuration has been
chosen such that the neutron is ejected at extreme backward angles, and the two protons at extreme
forward angles. The measurement was carried out at GeV laboratory energies. The experimental data
\cite{komarov} have been analyzed in Ref.~\cite{haidenbauer} using first and second order processes in
the NN $t$-matrix including a $\Delta$-isobar mechanism. Within our nonrelativistic toy model for three
bosons we are of course unable to analyze the data. However, within our model we can give a clear answer
whether higher order rescattering processes are essential in this reaction. In \cite{komarov,haidenbauer}
the data are integrated over a small interval of the relative $pp$ energy between 0 and 3~MeV, and
averaged over the neutron c.m. angle in the interval between 172$^o$ and 180$^o$. In our qualitative
study we fix the c.m. angle of one particle at 180$^o$, but integrate over the relative energy
$E_{pp}=p^2/m$ of the two other particles between 0 and 3~MeV. Thus, we evaluate the cross section
\begin{equation}
\frac{d\sigma}{d{\Omega_q}}= (2\pi)^4 \left( \frac{2}{3} \right)^2 \frac{m^2}{q_0} \int_0^{\sqrt{mE_{pp}}}
dp \; p^2 \; q \int d{\hat p} |U_0 (p,x_p,x^{q_0}_{pq},x_q=-1,q,q_0|^2.
\label{eq:5.1}
\end{equation}
Since we choose the z-axis parallel to ${\bf {\hat q}_0}$, and ${\bf {\hat q}}$ is antiparallel to
${\bf {\hat q}_0}$, the $\varphi_p$ dependence is directly given by $x^{q_0}_{pq} = \cos \varphi_p$. 
Our calculations are carried out for projectile laboratory energies between 0.2 and 1~GeV, and are
displayed in Fig.~\ref{fig12}. Here we compare different low orders in the two-body $t$-matrix with the full
solution of the Faddeev equation.  
We notice that all our calculations exhibit a smooth fall-off as function of the projectile energy.
This behavior is present in the data of Ref.~\cite{komarov}. None of our calculations shows a dip
structure around 0.7~GeV as indicated for some of the calculations in Ref.~\cite{haidenbauer}.
The reason may be that our calculation is carried out in three dimension, i.e. all partial waves are
included exactly, where as in Ref.~\cite{haidenbauer} only the lowest partial waves are considered.
 At low projectile energies rescattering
terms of higher order still give considerable contributions to the cross section. At 1~GeV the first
order calculation is an order of magnitude smaller than the result of the full calculation. It is
interesting to notice that at 1~GeV the contribution from the first rescattering is relatively small,and one needs to go to the 3rd order in $t$ to come close to the full result for this particular
break-up configuration.

In addition to the specific break-up configuration described above, a measurement of the extreme backward
scattering elastic $pd$ cross section has been investigated in \cite{haidenbauer}. Instead of a forward
scattered $pp$ pair with very small relative energy, one now has a forward going deuteron. 
This situation corresponds to elastic scattering from a deuteron at backward angle. To investigate the
influence of rescattering for this reaction we plot in Fig.~\ref{fig13} the backward angle of the elastic
cross section at energies from 0.2 to 1~GeV, and compare the result of the full Faddeev calculation with
calculations based on low orders in the multiple scattering series. The figure shows that 
the first order calculation is insufficient over the entire energy regime considered, except of course
for the crossing point around 0.5~GeV. The first rescattering contribution (2nd order calculation),
though close at 0.2~GeV, is insufficient below roughly 0.9~GeV. The figure also shows that at about
0.9~GeV the relative magnitude of contributions from the second and higher rescattering become small.
Thus, we conclude that at 1~GeV one needs at least one rescattering  to be in the vicinity of the full
result for the elastic cross section at the backward angle.

\section{Summary and Conclusions}

In this study we perform  fully converged
Faddeev calculations for three identical bosons interacting by non-separable
forces in the intermediate energy range between about 0.2 and 1.0 GeV. To the best
of our knowledge these are the first calculations of this kind. 
The key point is to neglect the
partial wave decomposition  generally used at low  energies and to work directly with momentum
vectors. Thus all partial waves are exactly included. Important is the suitable  choice of variables.
Besides the two magnitudes of the two relative Jacobi
momenta ${\bf p}$ and ${\bf q}$ we choose the angles between the vectors ${\bf p}$ and ${\bf q_0}$ 
and between ${\bf q}$ and ${\bf q_0}$, where ${\bf q_0}$ is the projectile  momentum.
The  fifth variable is the angle between the two planes spanned by ${\bf p}$, ${\bf q_0}$ 
and ${\bf q}$, ${\bf q_0}$.  In the technical piece of the work we introduce for the 
first time a spline based integration of the moving logarithmic singularities, which is a
very valuable alternative to procedures used so far. The numerical  results are 
converging as documented in Section III. In Section V we show elastic and 
inelastic (break-up) cross
sections in the above mentioned intermediate energy range.
We focus on the question how many orders of rescattering beyond the often used
first order calculation in the two-body $t$-matrix are needed to come close to the full Faddeev
result. We find that in nearly all cases studied processes of at least 2nd and 3rd order
rescattering are required. Whether this  will be also required in performing calculations
with realistic dynamical inputs has to be seen in the future. 

In one case we can make  first contact to a result based on the NN forces
AV18 and Bonn B, which are considered to be realistic in the sense that they describe all NN
data below 350 MeV extremely well. This was the semi-exclusive cross section at 495 MeV
 evaluated in first order in the NN t-matrix.
Of course, at that energy AV18 and Bonn-B are at the upper limit of their applicability. 
Despite our simple two-body model force, a superposition of two Yukawa interactions, 
one attractive the
other repulsive, our results turns out to be within about 20\% to the calculation based
on the realistic  models. 
This shows that our investigations might allow some conclusions about results
based on present and future models with more dynamical inputs.

As a first example for considering data in the light of our toy model
we study the extreme backward elastic dN scattering over the  energy range from 0.2 to 1.0 GeV.
We find  that first order results in the two-body $t$-matrix are totally insufficient and
only around 1 GeV the first order rescattering comes close to the full result. 
Parallel to those data in elastic
scattering in \cite{komarov,haidenbauer} the complete  break up process d(p,n)pp has also been
investigated. Here the neutron was ejected antiparallel to the beam direction and the
two protons at extreme forward angles with a very small  relative energy. Again we study
the significance of rescattering processes, and find that for this particular break-up  configuration two
rescatterings are necessary to get close to the result of the full Faddeev calculation.

In conclusion we can say that the three-body Faddeev equations can be safely solved 
at intermediate  energies using directly momentum vectors. Calculations based on partial
wave decomposition would be hardly feasible at these energies.

Further studies scanning the complete three-body phase space for the  total break-up are underway.
This  may be important in order to shed light on previous theoretical analysis of p(d,ppn)
reactions which  relied on low order reaction mechanisms.

Based on our current experience  it appears that if low order rescattering processes
 will turn out to be sufficient for
certain phase space regions, realistic calculations including spin and isospin will be
feasible, even including three-body forces. A first step evaluating the d(p,n)pp break-up
cross section in first  order with a currently  used two-pion exchange three-nucleon 
force model is already
under way \cite{imampriv}. What is badly needed now are realistic models for nuclear forces in the
intermediate  energy regime we study. This paper allows us to conclude that it
will be feasible to extend the calculations to realistic dynamics.

In Ref.~\cite{Nd3d} the effect of relativistic kinematics was studied in the
(p,n) charge exchange reaction on deuterium between 0.1 and 0.5~GeV in a first
order Faddeev calculation. This work concluded that the effects due to relativistic
kinematics are quite visible at 0.5~GeV, specifically in the location of the
 position of the
QFS peak, which is purely determined by kinematics. Therefore, we should expect
that the relativistic kinematics will influence our results, especially at energies
larger than 0.5~GeV. Of course, there are other dynamical relativistic effects. For
energies below $\sim$0.25~GeV those relativistic effects were studied in
neutron-deuteron elastic scattering in Ref.~\cite{WitKam}. There it was found that
the combination of relativistic effects consistently incorporated is negligible
below 0.1~GeV, and manifests itself at 0.25~GeV mostly at large scattering angles.
What happens at the energies we considered when relativistic effects (kinematical
and dynamical ones) are
incorporated is uncharted territory so far, and we want to refrain from
speculation. Work in this direction is however underway.

%-------------------------------------------------------------------------------

\vfill

\section*{Acknowledgments}
This work was performed in part under the
auspices of the U.~S.  Department of Energy under contract
No. DE-FG02-93ER40756 with Ohio University. We thank 
the National Energy Research Supercomputer Center (NERSC) for the use of
their facilities. We thank I.~Fachruddin for allowing us to use his results for comparison with ours. One
of us (W.G.) would like to thank K.~Miyagawa for fruitful collaboration on the spline based integration
of the logarithmic singularities.

%-------------------------------------------------------------------------------
\newpage

\appendix

\section{The $\varphi''$ integration}
\label{appendixa}

According to Eqs.~(\ref{eq:2.19}), (\ref{eq:2.18}), and (\ref{eq:2.20})
the $\varphi''$-integration 
for fixed $p$, $q$, $x_{p}$, $x_{q}$, $x^{q_{0}}_{pq}$, $q''$, and $x''$ 
can be written as
\begin{eqnarray}
I(\varphi_{q_{0}},\varphi_{p})
=\int^{2\pi}_{0}F[\cos(\varphi''-\varphi_{q_{0}})]G[\cos(\varphi''-\varphi_{p})]d\varphi''.
\label{A1}
\end{eqnarray}
where the $F$ and $G$ are known functions from $\hat{t}_{s}$ and
$\hat{T}$. The substitution  $\varphi'=\varphi''-\varphi_{q_{0}}$ leads to
\begin{eqnarray}
I(\varphi_{q_{0}},\varphi_{p})
&=&\int^{2\pi}_{0}
F[\cos\varphi']G[\cos(\varphi'-(\varphi_{p}-\varphi_{q_{0}}))]d\varphi'
\nonumber \\
&\equiv &I(\varphi_{q_{0}}-\varphi_{p}).
\label{A2}
\end{eqnarray}
Moreover, splitting this integral as
\begin{eqnarray}
I(\varphi_{q_{0}}-\varphi_{p})&=&
\int^{\pi}_{0}F[\cos\varphi']G[\cos(\varphi' -(\varphi_{p}-\varphi_{q_{0}}))]d\varphi'
\nonumber \\
&+& \int^{2\pi}_{\pi}F[\cos\varphi']G[\cos(\varphi'-(\varphi_{p}-\varphi_{q_{0}}))]d\varphi'
\label{A3}
\end{eqnarray}
and substituting 
$\varphi'=2\pi-\varphi''$ in the second integral, one obtains
\begin{equation}
I(\varphi_{p}-\varphi_{q_{0}})=
\int^{\pi}_{0} F[\cos\varphi''] \Biggl ( G[\cos(\varphi''-(\varphi_{p}-\varphi_{q_{0}}))]
+G[\cos(\varphi''+(\varphi_{p}-\varphi_{q_{0}}))] \Biggr )
d\varphi'' \equiv I (|\varphi_{p}-\varphi_{q_{0}}|).
\label{A4}
\end{equation}

Consequently, the result for the $\varphi$-integration in Eq.~(\ref{A2}) or  Eq.~(\ref{A4}) does not depend
on the choice of the sign in $\sin (\varphi_{p}-\varphi_{q_{0}}) = \pm \sqrt{1-\cos^2
(\varphi_{p}-\varphi_{q_{0}})}$. Only $\cos (\varphi_{p}-\varphi_{q_{0}})$ is fixed by Eq.~(\ref{eq:2.21}),
and has to be known.

Since the integral  $I$ in Eq.~(\ref{A4}) depends only on the difference of the angles
$(\varphi_{p}-\varphi_{q_{0}})$, one can choose $\varphi_{q_{0}}$ arbitrarily, and thus $\sin
\varphi_{q_{0}}$ and $\cos \varphi_{q_{0}}$ required in Eq.~(\ref{eq:2.18}). 
Moreover, the trivial identities
\begin{eqnarray}
\cos\varphi_{p} &=&
\cos\varphi_{q_{0}}\cos(\varphi_{p}-\varphi_{q_{0}}) -
\sin\varphi_{q_{0}}\sin(\varphi_{p}-\varphi_{q_{0}}), \nonumber \\
\sin\varphi_{p} &=&
\sin\varphi_{q_{0}}\cos(\varphi_{p}-\varphi_{q_{0}}) +
\cos\varphi_{q_{0}}\sin(\varphi_{p}-\varphi_{q_{0}}).
\label{A5}
\end{eqnarray}
are the input for $\cos (\varphi_{p}-\varphi'')$ needed in Eq.~(\ref{eq:2.20}). The arbitrary choice
of $\varphi_{q_{0}}$ is a good check for the numerical correctness of the choice of variables, and
we carried out those tests.

%------------------------------------------------------------------------------
%\newpage

\clearpage
%%%%%%%%%%%%%%%%%%%%%%%%%%%%%%%%%%%%%%%%%%%%%%%%%%%%%%

\begin{table}
\begin{tabular}{|c|ccccccc|ccc|c|} \hline
$E_{lab} [GeV]$ & $p$ & $x_{p}$ & $x^{q_0}_{pq}$ & $x_{q}$ & $q,q''$ & $x''$
& $\varphi''$ & $\sigma_{opt} $ [mb] & $\sigma_{el}$ [mb] &
$\sigma_{br}$ [mb] & $\sigma_{el}$ + $\sigma_{br}$ [mb] \\ \hline

0.01 &  49 &  4 &  4 &  4 & 49 &  4 &  4 & 1913.48 & 1799.08 & 67.81 & 1866.89  \\
     &  49 &  8 &  8 &  8 & 49 &  8 &  8 & 1886.84 & 1807.50 & 70.14 & 1877.64\\
     &  49 & 12 & 12 & 12 & 49 & 12 & 12 & 1904.99 & 1820.77 & 73.75 & 1894.52 \\
     &  49 & 16 & 16 & 16 & 49 & 16 & 16 & 1903.22 & 1820.46 & 73.20 & 1893.66 \\ \hline
\hline
0.1  &  49 & 12 & 12 & 12 & 49 & 12 & 12 & 335.57 & 259.95 & 83.10 & 343.05 \\
     &  49 & 16 & 16 & 16 & 49 & 16 & 16 & 343.17 & 265.83  &75.84  & 341.67  \\
     &  49 & 23 & 23 & 16 & 49 & 16 & 20 & 344.34 & 270.05 & 76.23 & 346.28\\
     &  49 & 23 & 23 & 24 & 49 & 24 & 20 & 346.16 & 272.04 & 76.55 & 348.59 \\  \hline
\hline
0.5  & 49  & 12 & 12 & 12 & 49 & 12 & 12 & 40.17 & 12.05 & 66.32 & 78.37\\
     & 49  & 16 & 16 & 16 & 49 & 16 & 16 & 65.62 & 47.76 & 32.73 & 80.49\\
     & 49  & 20 & 20 & 16 & 49 & 16 & 20 & 65.93 & 47.61 & 38.16 & 86.22\\
     & 49  & 20 & 16 & 20 & 49 & 20 & 16 & 85.19 & 61.16 & 28.84 & 90.00\\  \hline
     & 49  & 20 & 20 & 20 & 49 & 20 & 20 & 85.71 & 61.30 & 29.86 & 91.19\\
     & 49  & 24 & 20 & 20 & 49 & 20 & 20 & 85.72 & 61.24 & 30.56 & 91.80\\
     & 49  & 20 & 20 & 24 & 49 & 24 & 20 & 102.17 & 64.96 & 33.74& 98.70 \\
     & 49  & 23 & 23 & 24 & 49 & 24 & 20 & 110.35 &64.28  & 36.42& 100.70\\  \hline
\hline
\end{tabular}
\caption{The total elastic and break-up cross sections together with the total cross section
extracted via the optical theorem calculated from a Malfliet-Tjon type potential at two
selected energies (0.01 and 0.5~ GeV) as function of the grid points. 
The double prime quantities are the integration variables. The calculations are
carried out in the coordinate system in which $q_0$ is aligned parallel to the
z-axis.
}
\label{table-2}
\end{table}

\begin{table}
\begin{tabular}{|c|cc|ccc|cc|}
\hline
$E_{lab}$ [GeV] &  $\sigma^{q_{0}}_{el}$ [mb] & $\sigma^{q}_{el}$ [mb] &
$\sigma^{q_{0}}_{br}$ [mb] & $\sigma^{q}_{br}$ [mb] & $\sigma^{p}_{br}$ [mb] &
$\sigma^{q_{0}}_{opt}$ [mb]  &  $\sigma^{q}_{opt}$ [mb]  \\ \hline
0.003& 2561.74 & 2561.14 & 0.0   & 0.0   &  0.0  & 2562.65 & 2562.65 \\ \hline
0.01 & 1820.46 & 1820.51 & 73.20 & 73.55 & 73.13 & 1903.22 & 1902.56 \\ \hline
0.1  & 272.04  &  272.20 & 76.55 & 75.18 & 75.08 &  346.16 & 346.16 \\ \hline
0.5  &  64.28  &  64.61  & 36.42 & 36.39 & 35.55 &  110.35 &  110.35 \\ \hline
1.0  &  21.90  &  21.90 &  23.44 & 23.46 & 23.40 &   49.59 &   49.59 \\ \hline
\end{tabular}
\caption{The total elastic cross section, total breakup cross
section and total cross section extracted via the optical theorem
 calculated in different coordinate systems at selected
energies. The choice of coordinate system, i.e. which vector is aligned parallel to the
z-axis, is indicated by the superscripts $q_0$, $q$ and
$p$. }
\label{table-3}
\end{table}

\begin{table}
\begin{tabular}{|c|c|c|c|} \hline
 $E_{lab}$ [GeV]  & $\theta_{cm} $ [deg]   &
 $\frac{d\sigma^{el}}{d\Omega_{cm}}|_{T}$ [mb] &
 $\frac{d\sigma^{el}}{d\Omega_{cm}}|_{T'}$ [mb] \\\hline
 0.01      &        0.0      &          537.536      &           537.536 \\
           &       21.8      &          420.036      &           420.036  \\
           &       62.1      &           70.726      &            70.725  \\
           &       93.4      &           38.289      &            38.289  \\
           &       151.5     &          227.899      &           227.899   \\ \hline
 0.2       &        0.0      &          676.821      &          676.821  \\
           &       21.8      &          148.880      &          148.880   \\
           &       62.1      &            0.363      &            0.363  \\
           &       93.4      &            0.223      &            0.223  \\
           &       151.5     &            0.010      &            0.010  \\ \hline
 0.5       &        0.0      &         519.389      &          519.389  \\
           &       21.8      &          16.209      &           16.209  \\
           &       26.3      &           4.430      &            4.430  \\
           &       62.1      &           0.088      &            0.088  \\
           &       93.4      &           0.005      &            0.005  \\
           &       151.5     &           0.004      &            0.004  \\ \hline
 1.0       &        0.0      &    $3.903\times 10^{+2}$   & $3.903\times 10^{+2}$  \\
           &       21.8      &    $5.325\times 10^{-1}$  & $5.325\times 10^{-1}$  \\
           &       62.1      &    $4.072\times 10^{-4}$  & $4.072\times 10^{-4}$ \\
           &       93.4      &    $2.678\times 10^{-3}$  & $2.678\times 10^{-3}$  \\
           &       151.5     &    $3.705\times 10^{-4}$  & $3.703\times 10^{-4}$  \\
\hline
 \end{tabular}
\caption{\label{table4} 
The elastic differential cross sections at different energies 
for selected scattering angles. The cross section labeled by $T$ results from
the converged solution of the integral equation Eq.~(\ref{eq:2.19}). The column labeled $T'$
is calculated by reinserting the original solution into the Faddeev equation with
 $T'=tP + tG_{0}PT$. The calculations are based on a Malfliet-Tjon type potential as
described in the text. 
}
\label{table-4}
\end{table}

\vspace{5mm}

\begin{table}
\begin{tabular}{|c|cc|cc|}
\hline
$\varphi_{q_{0}}$ [rad] &  $\sigma^{q_{0}}_{el}$ [mb] & $\sigma^{q}_{el}$ [mb] &
$\sigma^{q_{0}}_{opt}$ [mb] & $\sigma^{q}_{opt}$ [mb] \\ \hline
0.0              & 2561.736 & 2561.138 & 2562.649  & 2562.649   \\ \hline
$\frac{\pi}{2}$  & 2560.885 & 2560.729 & 2562.496  & 2562.496   \\ \hline
$\pi$            & 2561.550 & 2561.206 & 2562.091  & 2562.091   \\ \hline
\end{tabular}
\caption{ The total elastic cross sections at $E_{lab}=3.0$ MeV calculated
for different values of the angle $\varphi_{q_{0}}$  with the $+$ 
sign of $\sin(\varphi_{p}-\varphi_{q_{0}})$. The calculations are carried out in two
different coordinate systems, characterized by the superscripts $q_0$ and $q$, which
indicate, which vector is chosen to be parallel to the z-axis.} 
\label{table-5}
\end{table}

\vspace{5mm}

\begin{table}
\begin{tabular}{|c|cc|cc|}
\hline
${\mathrm{sign}}[\sin(\varphi_{p}-\varphi_{q_{0}})]$ & 
 $\sigma^{q_{0}}_{el}$ [mb] & $\sigma^{q}_{el}$ [mb] &
$\sigma^{q_{0}}_{opt}$ [mb] & $\sigma^{q}_{opt}$ [mb]\\ \hline
+  & 2561.736 & 2561.138 & 2562.649 & 2562.649     \\ \hline
-  & 2559.674 & 2559.536 & 2560.091 & 2560.091    \\ \hline
\end{tabular}
\caption
{The total elastic cross sections at $E_{lab}=3.0$ MeV calculated
for different sign of $\sin(\varphi_{p}-\varphi_{q_{0}})$ where  $\varphi_{q_{0}}=0$.
The meaning of the superscripts is the same as in Table~\ref{table-5}.}
\label{table-6}
\end{table}

\vspace{5mm}
 
\begin{table}
\begin{tabular}{|c|c|c|c|c|}
\hline
 $V_{A}$ [MeV fm] & $\mu_{A}$ [${\mathrm{fm}}^{-1}$] & $V_{R}$[MeV fm]
  & $\mu_{R}$ [${\mathrm{fm}}^{-1}$] & $E_{d}$[MeV] \\ \hline
  -626.8932 & 1.550 & 1438.7228 & 3.11 & -2.2307  \\  \hline
  \end{tabular}
\caption
{The parameters and deuteron binding energy for the Malfliet-Tjon type
potential of our calculation. As conversion factor 
We use units such that $\hbar c$=197.3286~MeV fm = 1.}
\label{table-7}
\end{table}

\clearpage
%%%%%%%%%%%%%%%%%%%%%%%%%%%%%%%%%%%%%%%%%%%%%%%%%%%%%%

\noindent

\begin{figure}
\begin{center}
\includegraphics[width=8cm]{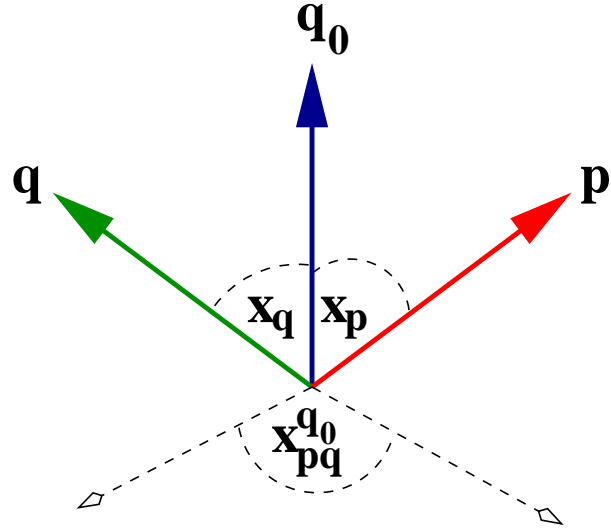}
\end{center}
\caption{ The geometry of three vectors ${\bf q_0}$, ${\bf q}$, and ${\bf p}$
relevant in the three-body scattering problem. The independent angle variables $x_q$, $x_p$, and
$x^{q_{0}}_{pq}$ as defined in Section~II are indicated. The dashed arrows represent 
the normal vectors (${\bf q_0} \times {\bf q}$) and (${\bf p} \times {\bf q_0}$).  
\label{fig1}}
\end{figure}

\begin{figure}
\begin{center}
\includegraphics[width=9cm]{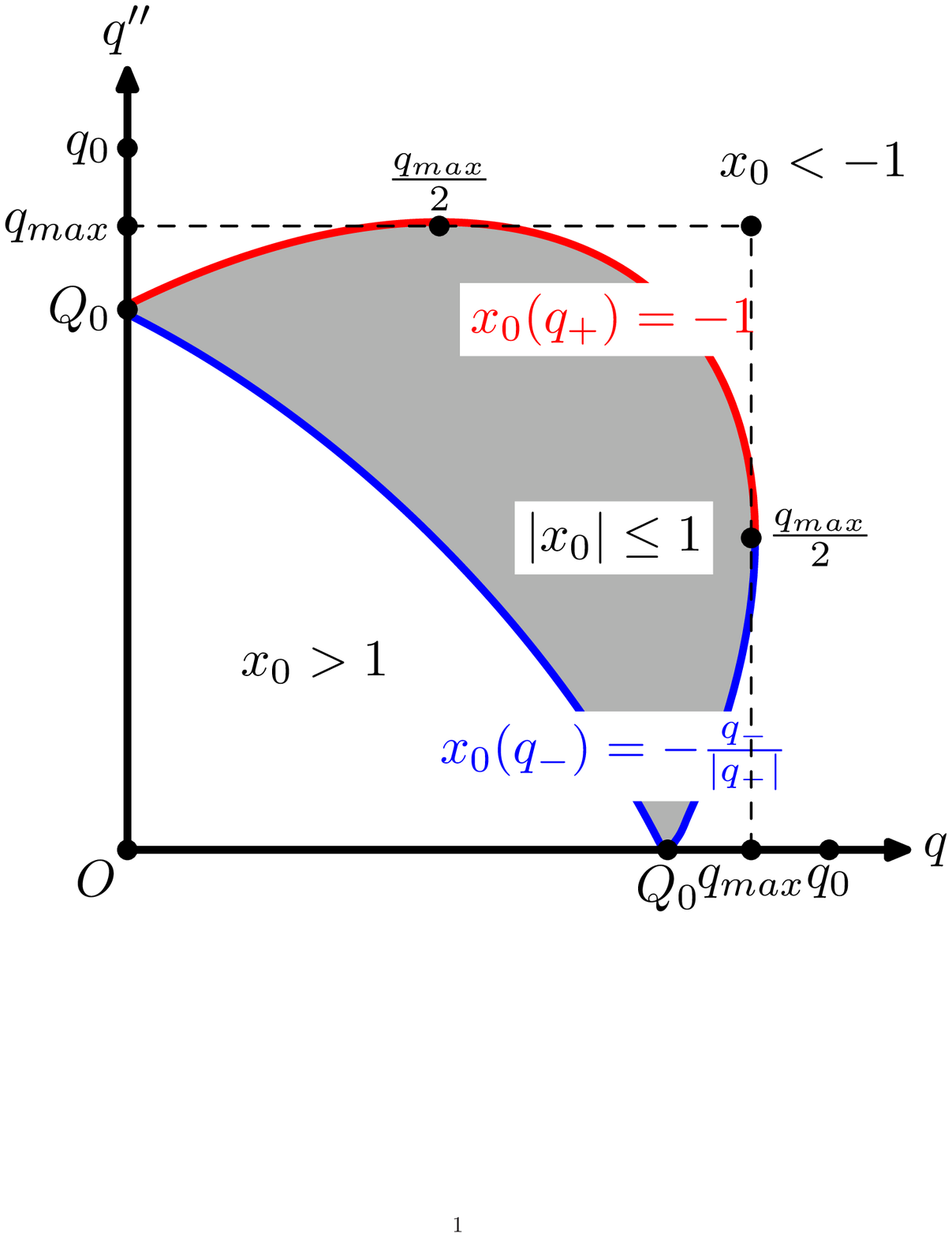}
\end{center}
\caption{ The region of singularities of the free three-particle propagator as function of
the momenta $q$ and $q''$.  The shaded area in the 
$q-q''$ plane indicates the region where $|x_{0}|\le 1$, i.e. the region where a pole in the 
$x''$-integration occurs. This region is
is enclosed by the  bounding curves $q_{+}$ and $q_{-}$, which contain the logarithmic singularity 
as function of $q''$ as given in
Eqs.~(\ref{eq:4.5}) and (\ref{eq:4.6}).
\label{fig2}}
\end{figure}

\begin{figure}
\begin{center}
\includegraphics[width=9cm]{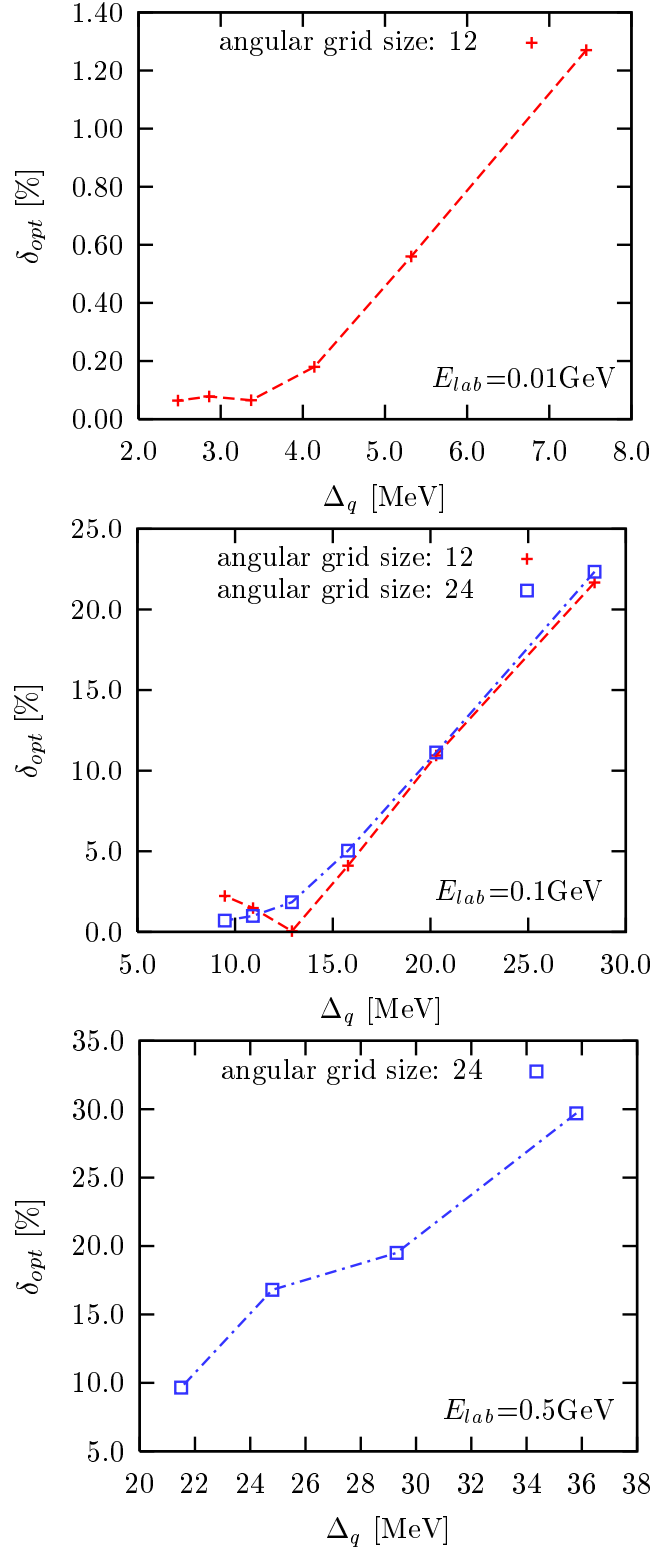}
\end{center}
\caption{The percent error in the optical theorem as function of the average
distance $\Delta_q$ of the integration grid points $q''$ in the interval
$(0,q_max)$ at the selected laboratory projectile energies $E_{lab}$=~0.01~GeV (top
panel), $E_{lab}$=~0.1~GeV (middle panel), and $E_{lab}$=~0.5~GeV (bottom panel).  
\label{fig3a}}
\end{figure}

\begin{figure}
\begin{center}
\includegraphics[width=12cm]{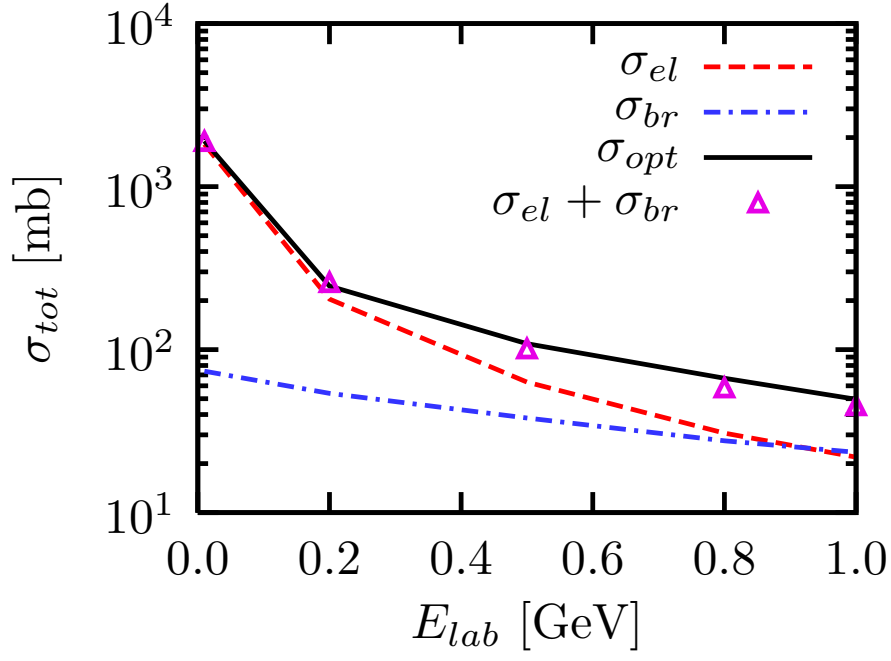}
\end{center}
\caption{ The total elastic cross section $\sigma_{el}$ (dashed line), 
the total break-up cross section $\sigma_{br}$ (dash-dotted line) and the
total cross section evaluated by the optical theorem $\sigma_{opt}$ (solid line)
given as function of the projectile laboratory energy. 
At selected energies where the calculations have been carried out
 the sum of the calculated total elastic and break-up 
cross section, $\sigma_{tot}=\sigma_{el}+\sigma_{br}$, is indicated by the open diamond.
The open diamonds coincide with the solid line according to the optical
theorem, Eq.~\ref{eq:3.22}, and the numerical values are given in Table~\ref{table-3}.
\label{fig3}}
\end{figure}

\begin{figure}
\begin{center}
\includegraphics[width=13cm,angle=-90]{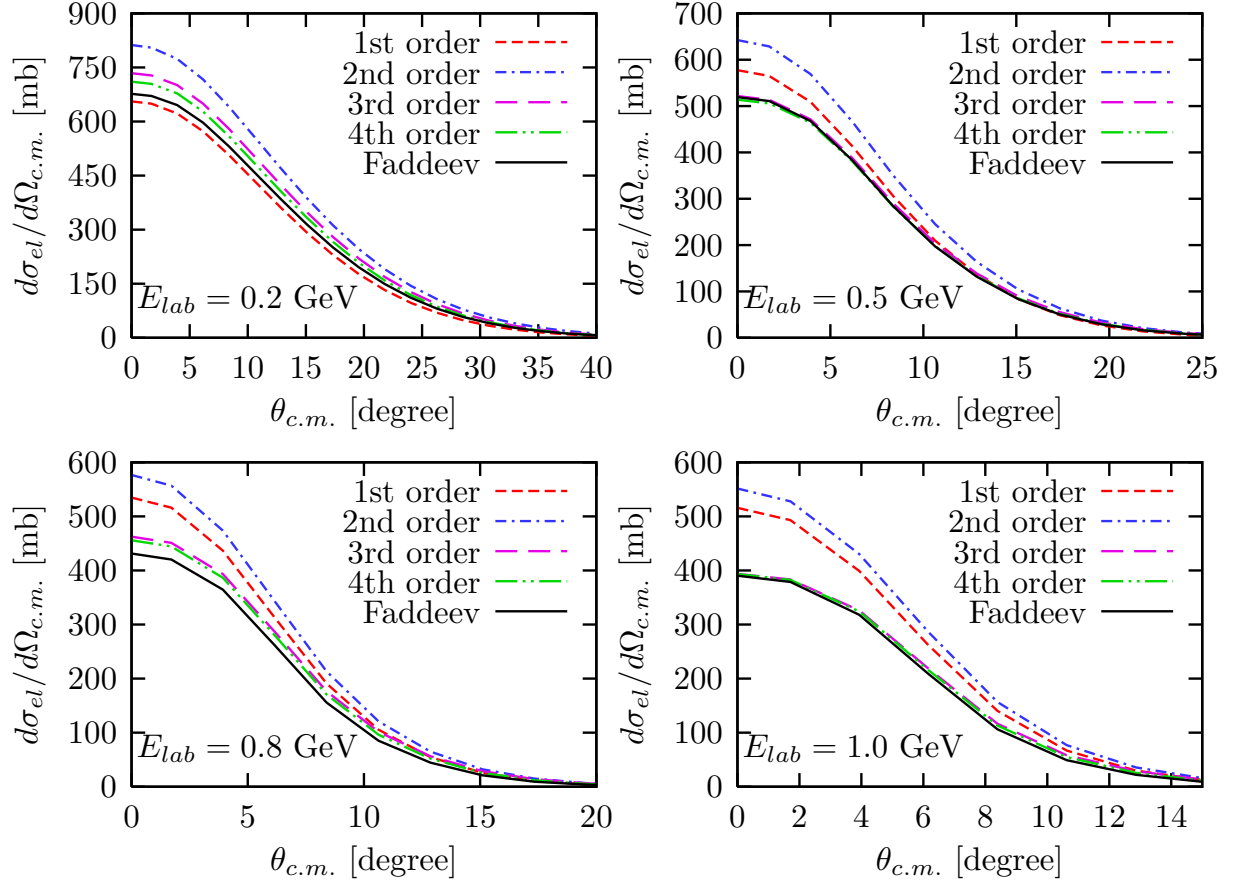}
\end{center}
\caption{The elastic differential cross section at 0.2~GeV, 0.5~GeV, 0.8~GeV, and 1.0~GeV
projectile energy as function of the laboratory scattering angle. In all cases the solid line
represent the full solution of the Faddeev equation. The other lines represent the successive
sum of different orders in the
multiple scattering series, short-dashed the first order, dash-dot adds up to 
the 2nd order, long-dashed to the 3rd
order, and dash-dot-dot to the 4th order contribution.
\label{fig4}}
\end{figure}

\begin{figure}
\begin{center}
\includegraphics[width=13cm,angle=-90]{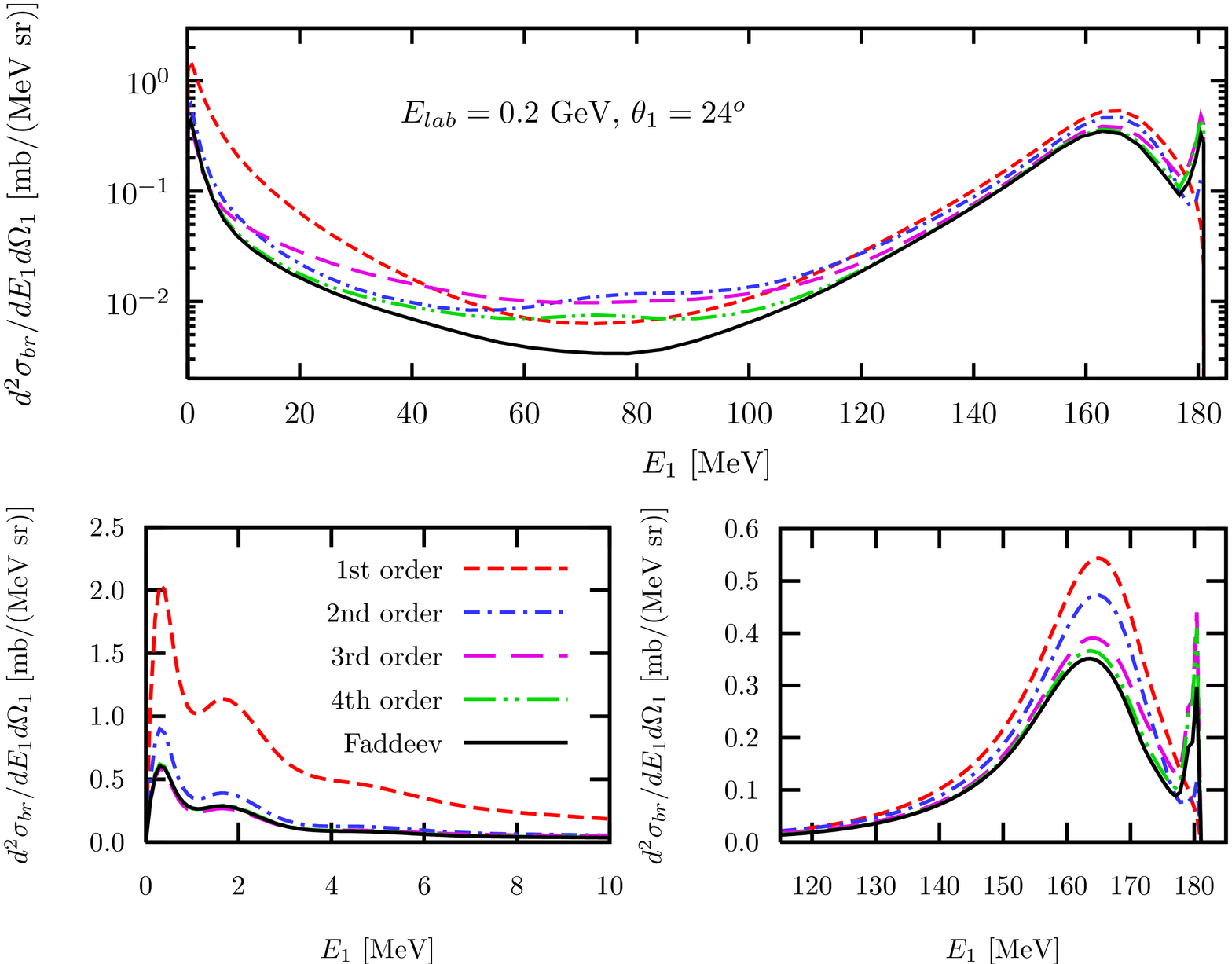}
\end{center}
\caption{ The semi-exclusive cross section at 0.2 GeV laboratory incident energy and at 24$^o$ emission
angle of
the emitted particle. The upper panel displays the entire energy range of the emitted particle, whereas
the two lower panels show only the low and high energies in a linear scale. The full solution of the
Faddeev equation is given by the solid line in all panels. The contribution of the lowest orders of the
multiple scattering series added up successively is given by the other curves as indicated in the legend
of the lower left panel.
\label{fig5}}
\end{figure}

\begin{figure}
\begin{center}
\includegraphics[width=13cm,angle=-90]{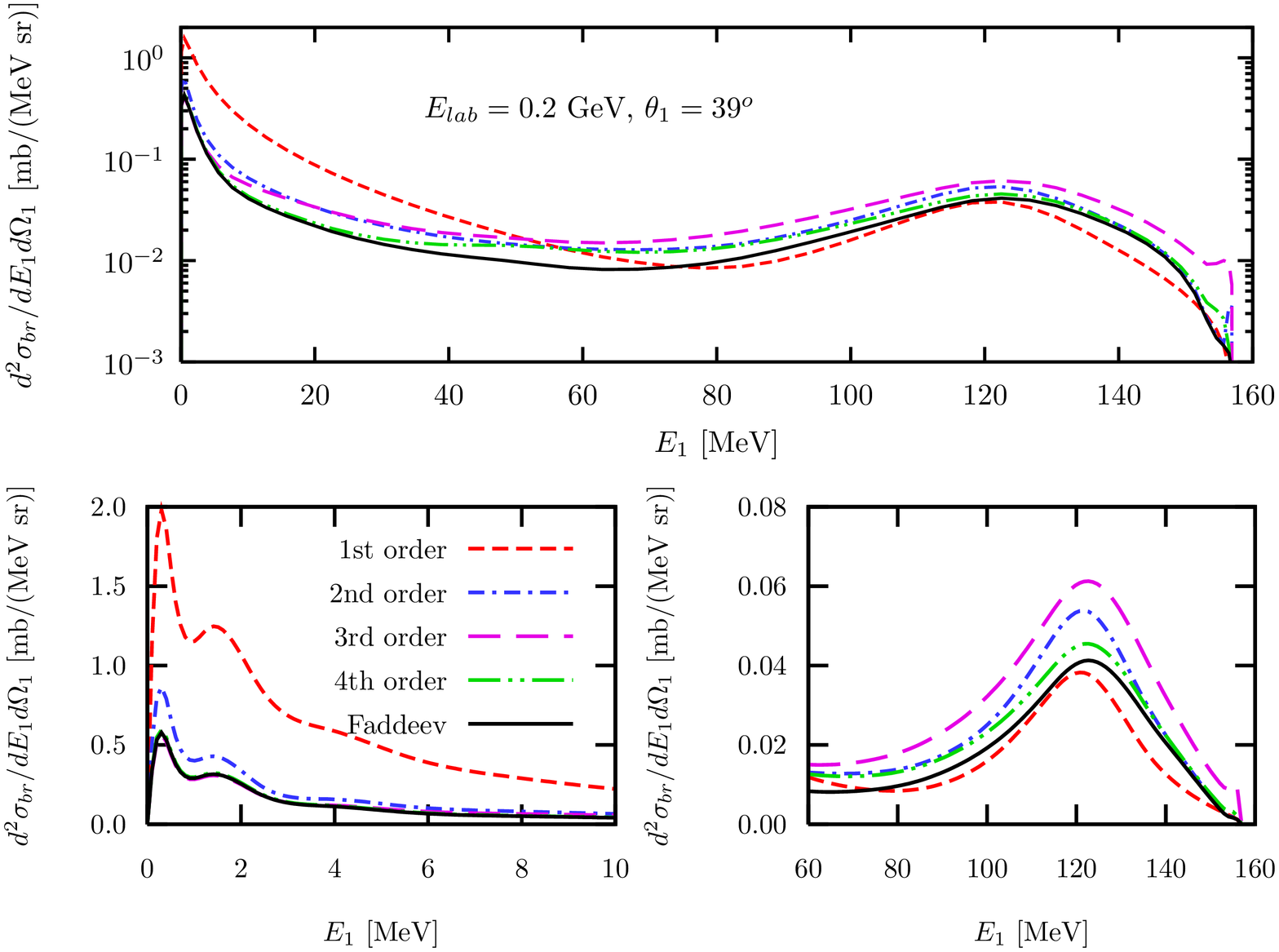}
\end{center}
\caption{ Same as Fig.~\ref{fig5} but for an angle of 39$^o$ of the emitted particle.
\label{fig6}}
\end{figure}

\begin{figure}
\begin{center}
\includegraphics[width=13cm,angle=-90]{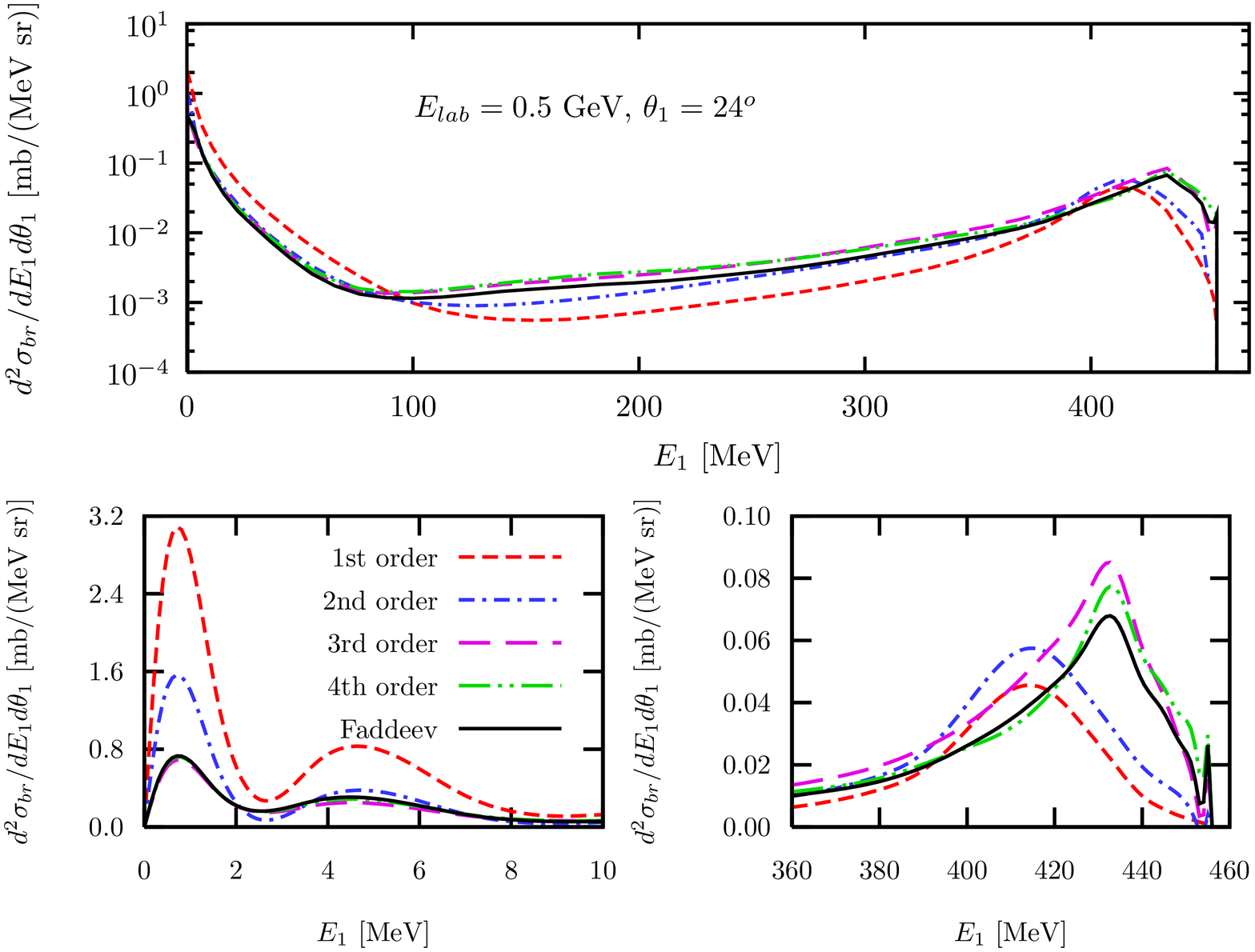}
\end{center}
\caption{ The semi-exclusive cross section at 0.5 GeV laboratory incident energy and at 24$^o$ angle of
the emitted particle. The upper panel displays the entire energy range of the emitted particle, whereas
the two lower panels show only the low and high energies in a linear scale. The full solution of the
Faddeev equation is given by the solid line in all panels. The contribution of the lowest orders of the
multiple scattering series added up successively is given by the other curves as indicated in the legend
of the lower left panel.
\label{fig7}}
\end{figure}

\begin{figure}
\begin{center}
\includegraphics[width=13cm,angle=-90]{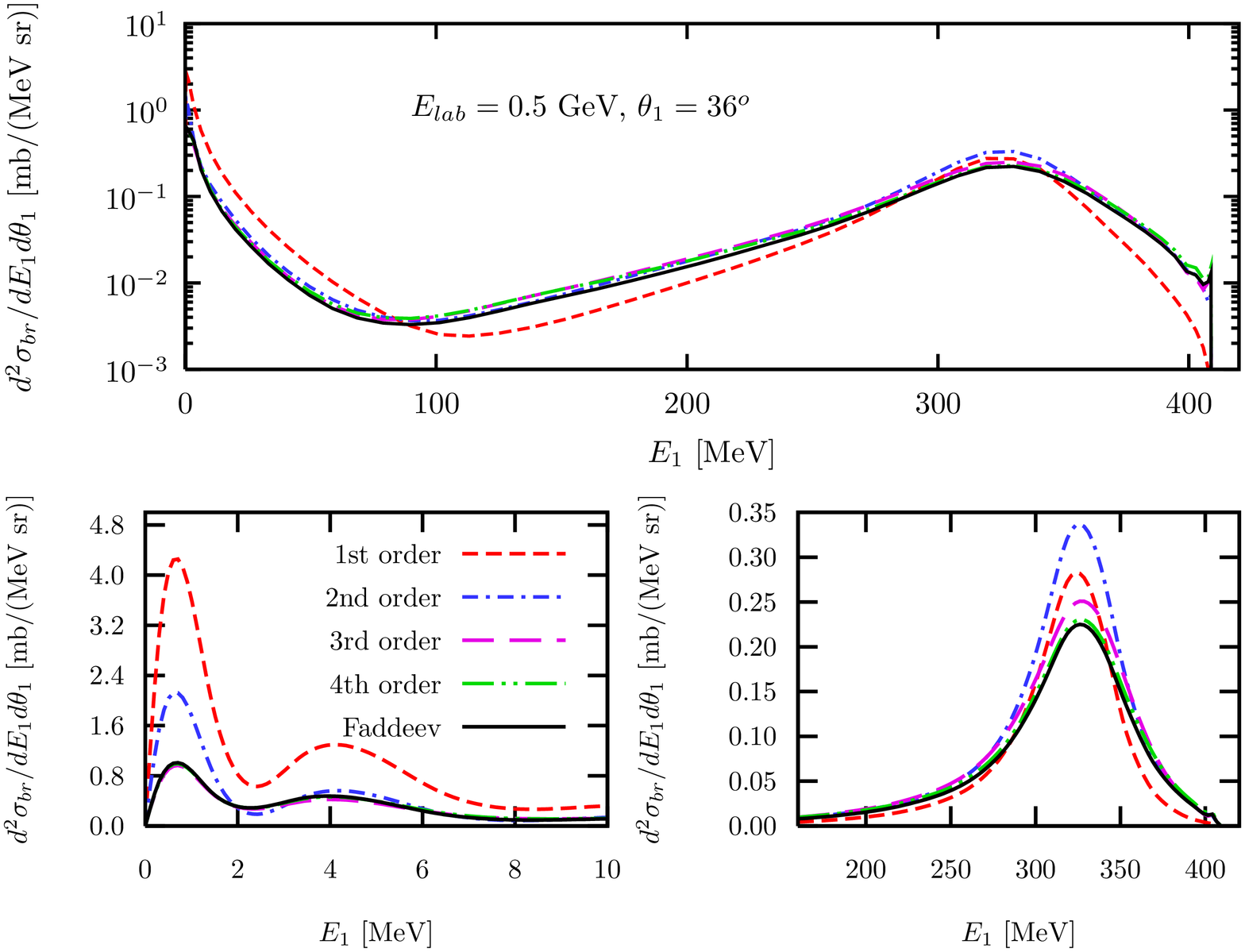}
\end{center}
\caption{ Same as Fig.~\ref{fig7} but for an angle of 36$^o$ of the emitted particle.
\label{fig8}}
\end{figure}

\begin{figure}
\begin{center}
\includegraphics[width=13cm,angle=-90]{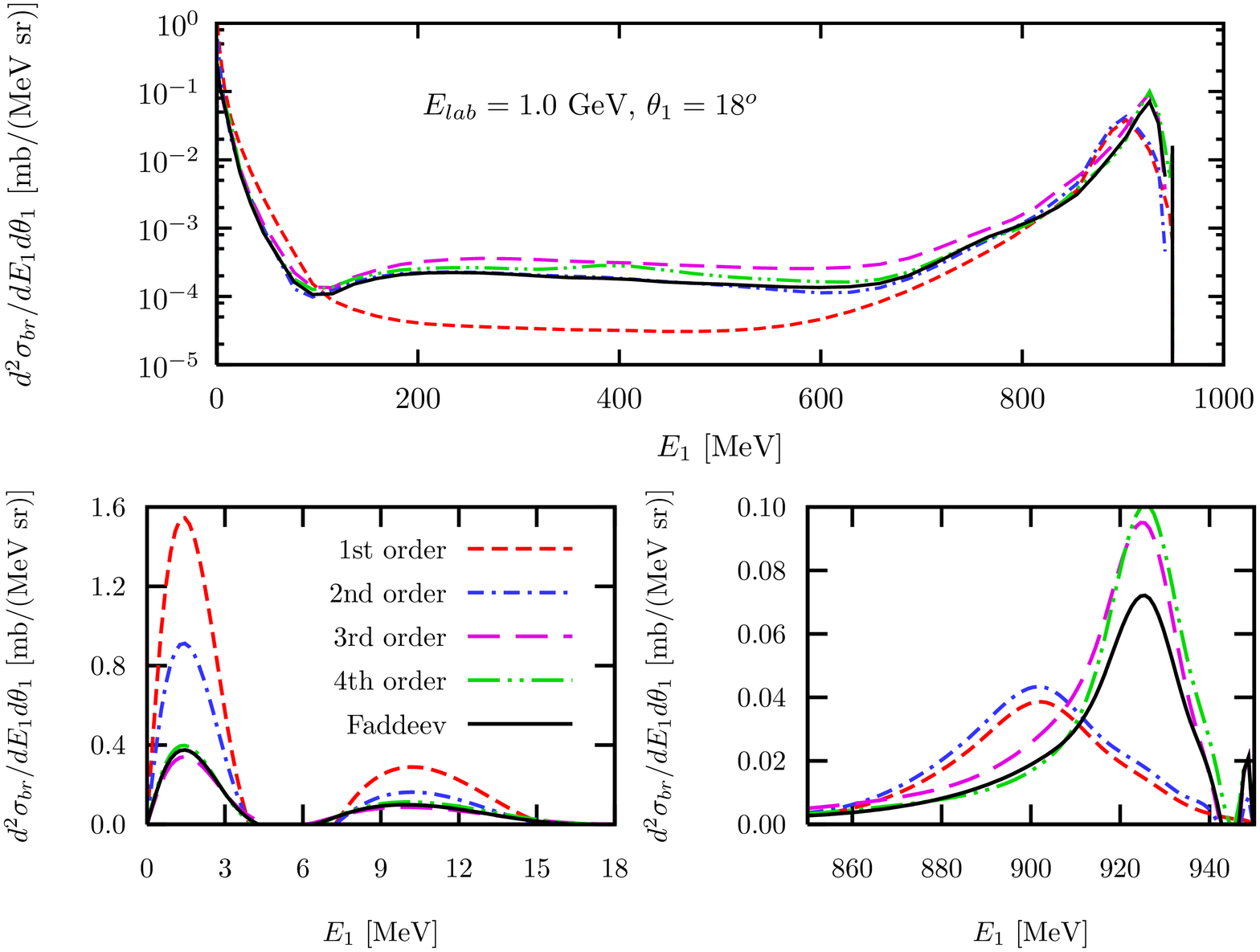}
\end{center}
\caption{ The semi-exclusive cross section at 1 GeV laboratory incident energy and at 18$^o$ angle of
the emitted particle. The upper panel displays the entire energy range of the emitted particle, whereas
the two lower panels show only the low and high energies in a linear scale. The full solution of the
Faddeev equation is given by the solid line in all panels. The contribution of the lowest orders of the
multiple scattering series added up successively is given by the other curves as indicated in the legend
of the lower left panel.
\label{fig9}}
\end{figure}

\begin{figure}
\begin{center}
\includegraphics[width=13cm,angle=-90]{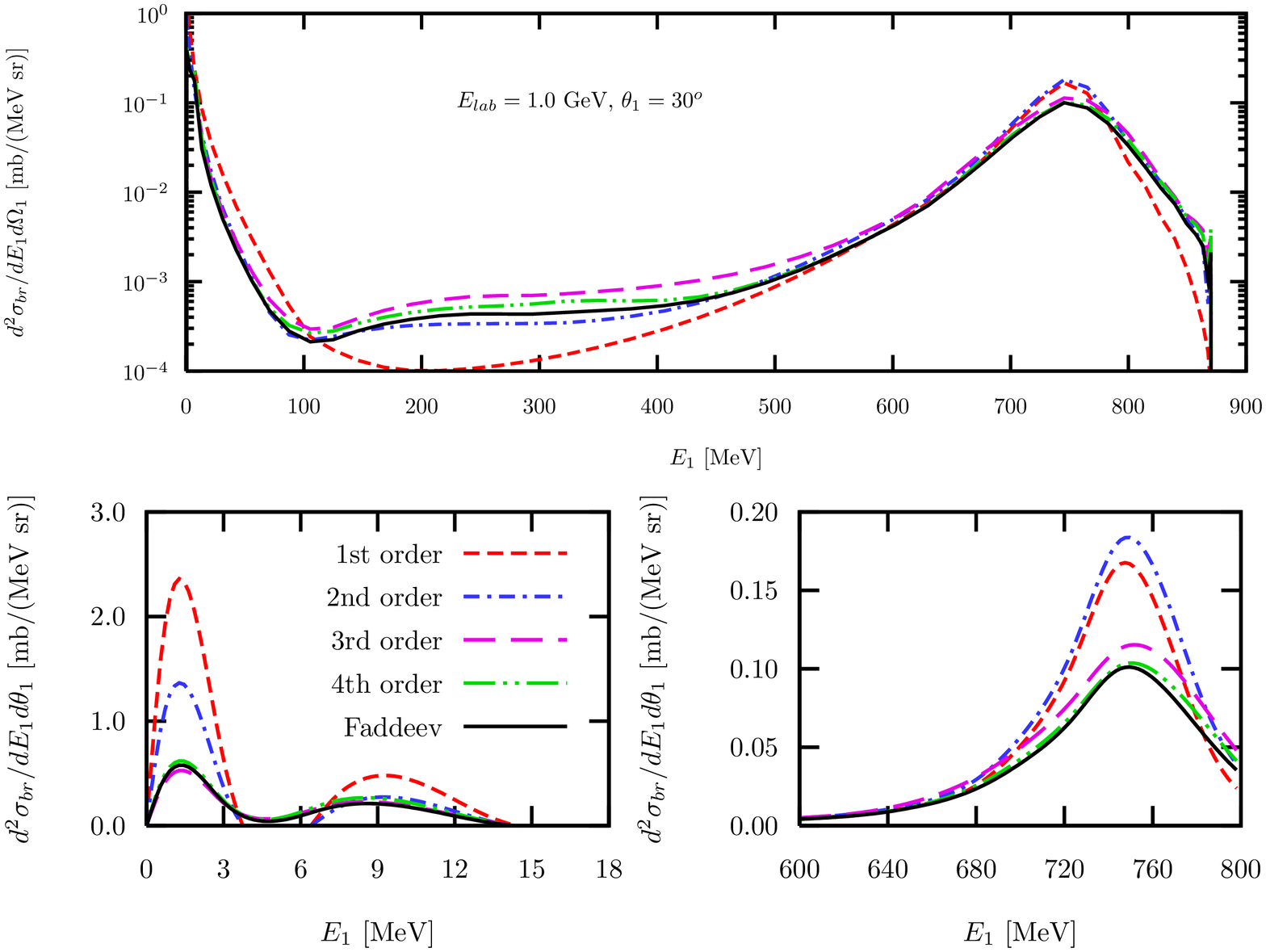}
\end{center}
\caption{ Same as Fig.~\ref{fig9} but for an angle of 30$^o$ of the emitted particle.
\label{fig10}}
\end{figure}

\begin{figure}
\begin{center}
\includegraphics[width=13cm]{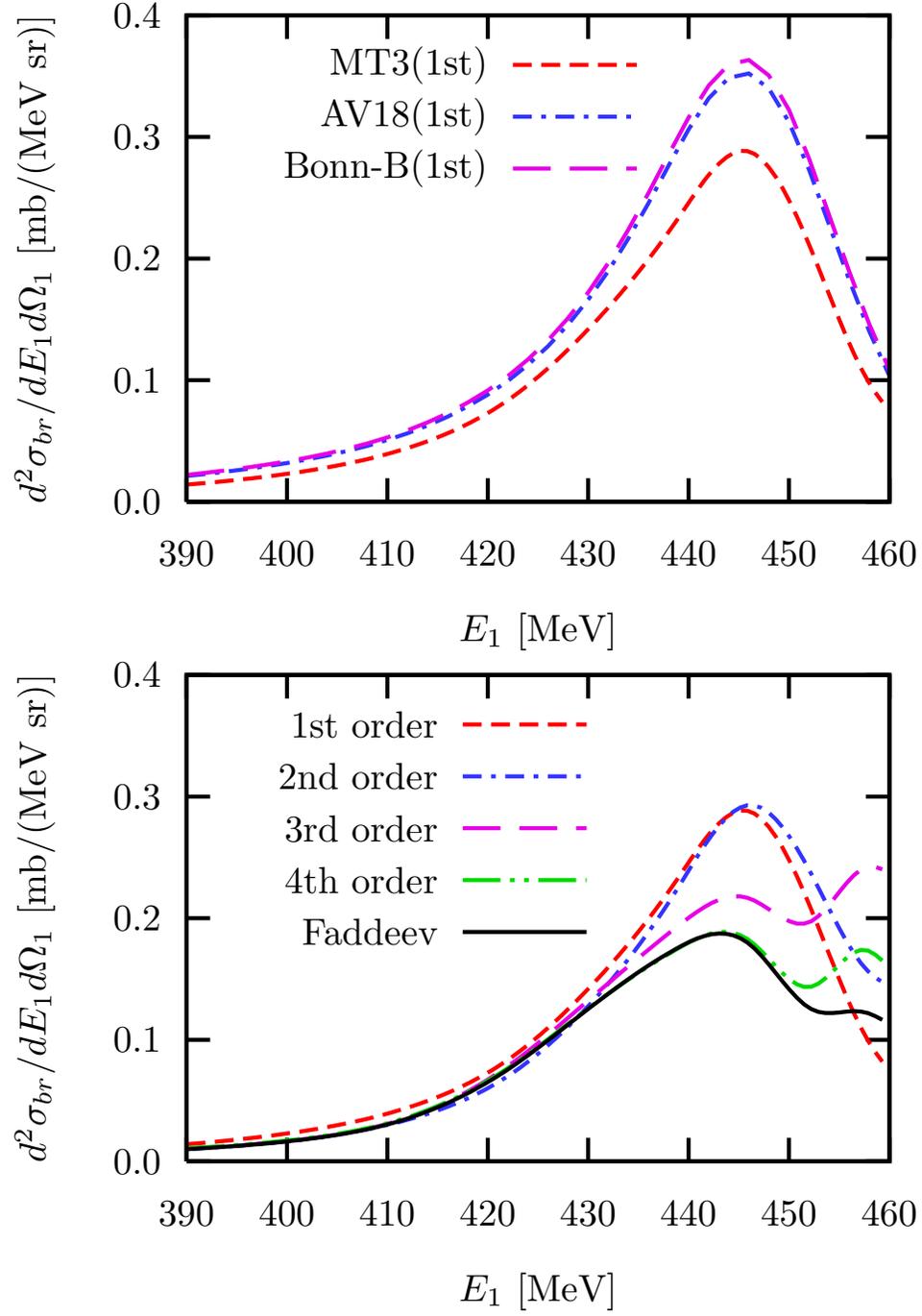}
\end{center}
\caption{ The semi-exclusive cross section at 495 MeV laboratory incident energy and at 18$^o$ angle of
the emitted particle. The upper panel displays the first order results obtained from the realistic
potentials AV18 \protect\cite{av18} (long dashed line) and Bonn-B \protect\cite{bonnb} (dashed-dotted line) together with
our calculation based on the scalar MT potential of Eq. \ref{eq:5.1} (dashed line). The lower panel 
displays again our first order calculation from the upper panel (dashed line), together with a
successive addition of the next three rescattering terms. The exact solution 
of the Faddeev equation is given by the solid line.
\label{fig11}}
\end{figure}

\begin{figure}
\begin{center}
\includegraphics[width=12cm]{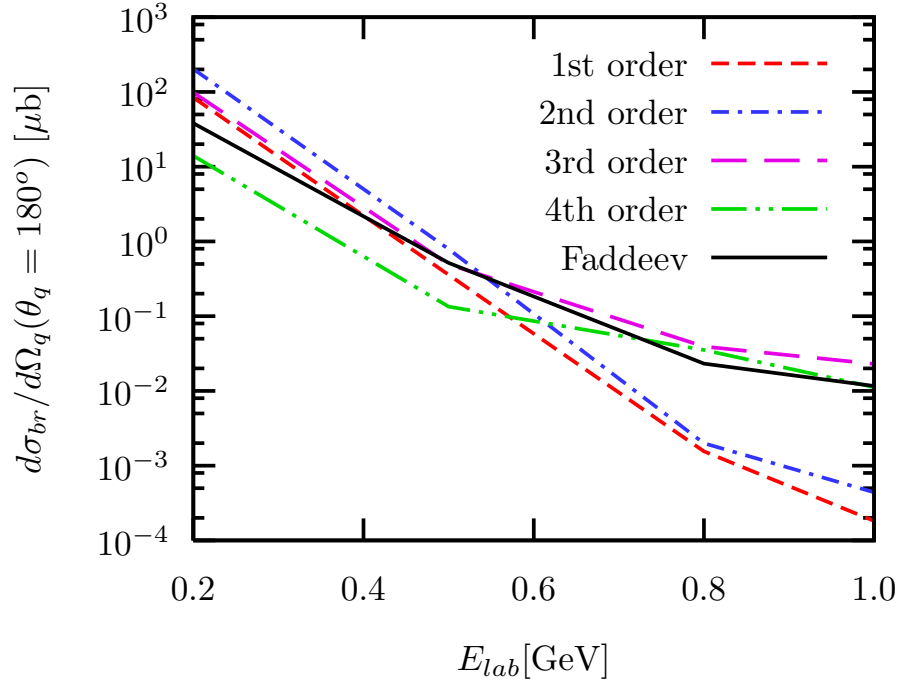}
\end{center}
\caption{
The cross section (c.m.) for the semi-exclusive break-up reaction in which two particles emerge in forward
direction with a relative energy between 0 and 3 MeV, and one particle is detected at backward angle as
function of the projectile laboratory energy. The result of the full Faddeev calculation is given by
the solid line and  is compared with
calculations based on the lowest orders of the multiple scattering series in $t$ added up successively 
as indicated in the legend.
\label{fig12}}
\end{figure}

\begin{figure}
\begin{center}
\includegraphics[width=12cm]{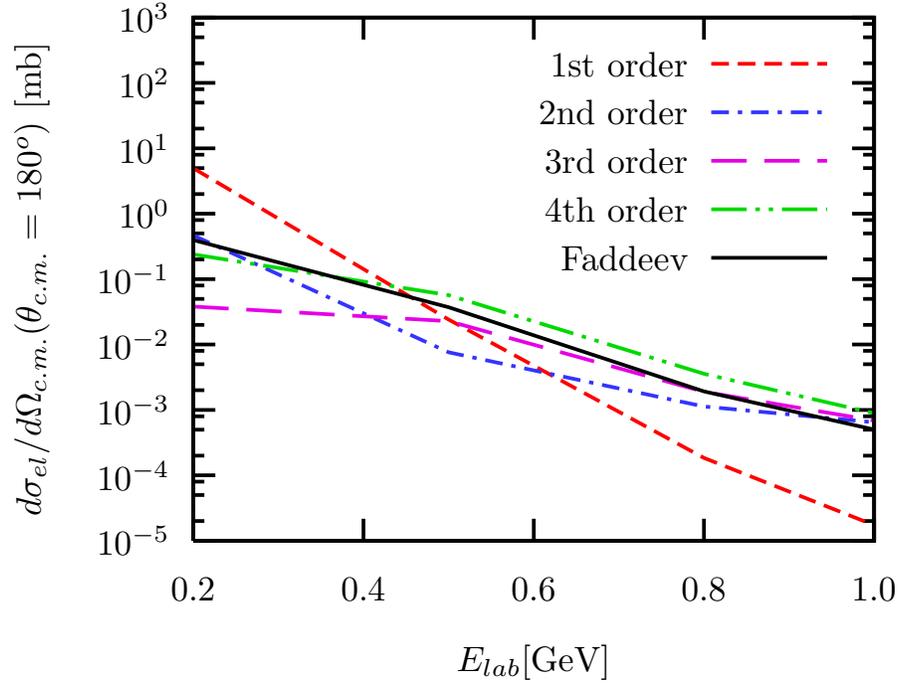}
\end{center}
\caption{
The elastic cross section (c.m.)  at backward angle as function of the projectile laboratory
energy. The result of the full Faddeev calculation is given by the solid line and compared with
calculations based  on the lowest orders of the multiple scattering series in $t$ added 
up successively as indicated in the legend. 
\label{fig13}}
\end{figure}

\end{document}